 \documentclass{aa}  
\usepackage{graphicx}
\usepackage[varg]{txfonts}
\usepackage{natbib}
\bibpunct{(}{)}{;}{a}{}{,} 

\usepackage{color}
\usepackage{subfigure}

\begin{document}

    \title{Squeezed between shells? On the origin of the Lupus~{\rm I} molecular cloud}
    
    \subtitle{APEX/LABOCA\thanks{The Atacama Pathfinder Experiment (APEX) is a 
collaboration between the Max-Planck-Institut für Radioastronomie (MPIfR), the European Southern Observatory (ESO), 
and the Onsala Space Observatory (OSO).}, {\it Herschel}\thanks{{\it Herschel} is an ESA space observatory with science instruments provided by European-led Principal Investigator consortia and with important participation from NASA.} and {\it Planck} observations}

   \author{B.~Gaczkowski\inst{1} \and T.~Preibisch\inst{1} \and T.~Stanke\inst{2} \and M.G.H.~Krause\inst{1,3,4}
\and A.~Burkert\inst{1,3} \and R.~Diehl\inst{3,4} \and K.~Fierlinger\inst{1,4} \and D.~Kroell\inst{1,3} \and  J.~Ngoumou\inst{1} \and V.~Roccatagliata\inst{1}}

   \institute{Universit\"ats-Sternwarte M\"unchen, 
        Ludwig-Maximilians-Universit\"at,
          Scheinerstr.~1, 81679 M\"unchen, Germany;
	  \email{bengac@usm.uni-muenchen.de}
\and
ESO, Karl-Schwarzschild-Strasse 2, 85748 Garching bei M\"unchen, Germany
\and
Max-Planck-Institut f\"ur extraterrestrische Physik, Postfach 1312, 85741 Garching, Germany
\and
Excellence Cluster Universe, Technische Universit\"at M\"unchen, Boltzmannstrasse 2, 85748 Garching, Germany}

 \titlerunning{Lupus~{\rm I}}

   \date{Received 13 May 2015;  Accepted 21 September 2015}

\abstract{The Lupus~{\rm I} cloud is found between the Upper-Scorpius (USco) and the Upper-Centaurus-Lupus (UCL) sub-groups of the
Scorpius-Centaurus OB-association, where the expanding USco H\,{\rm I} shell appears to
interact with a bubble currently driven by the winds of the remaining
B-stars of UCL.}
{We want to study how collisions
of large-scale interstellar gas flows form and influence new dense clouds in the ISM.}
{We performed LABOCA continuum sub-mm observations of Lupus~{\rm I} that provide for the first time a direct
view of the densest, coldest cloud clumps and cores at high angular resolution.
We complemented those by \textit{Herschel} and \textit{Planck} data from which we constructed column density
and temperature maps. From the \textit{Herschel} and LABOCA column
density maps we calculated PDFs to characterize the density structure of the cloud.}
{The northern part of Lupus~{\rm I} is found to have on average lower densities and higher temperatures
as well as no active star formation. The center-south part harbors dozens of pre-stellar cores
where density and temperature reach their maximum and minimum, respectively.
Our analysis of the column density PDFs from the \textit{Herschel} data show double peak profiles for all parts
of the cloud which we attribute to an external compression. In those parts with active star formation, the PDF shows a power-law tail at high densities.
The PDFs we calculated from our LABOCA data trace the denser parts of the cloud showing one peak and a power-law tail. With LABOCA we find 15 cores with masses
 between 0.07 and $1.71\,M_\odot$ and a total mass of $\approx8\,M_\odot$. The total gas and dust mass of the cloud is $\approx164\,M_\odot$ 
and hence $\sim5\%$ of the mass is in cores. From the \textit{Herschel} and \textit{Planck} data we find a total mass of $\approx174\,M_\odot$ and
$\approx171\,M_\odot$, respectively.}
{The position, orientation and elongated shape of Lupus~{\rm I}, the double peak PDFs and the population 
of pre-stellar and protostellar cores could be explained by the large-scale compression
from the advancing USco H\,{\rm I} shell and the UCL wind bubble.}

 
   \maketitle
%

\section{Introduction}

In the current picture of the dynamic interstellar medium (ISM), molecular cloud
formation is attributed to collisions of large-scale flows in the ISM \citep[see review by][
and references within]{mc-formation-ppvi-review-2014}. Such flows can be driven
by stellar feedback processes (e.g. UV-radiation and winds) and supernovae.
At the interface of the colliding flows, compression, cooling, and fragmentation of the diffuse atomic medium produces
cold sheets and filaments that later may become molecular and self-gravitating and dominate the
appearance of the ISM as observed today \citep[see review by][]{filaments-sf-ppvi-review-andre-2014}.
In this picture 
the fast formation (and dispersion) of molecular clouds and the often simultaneous onset 
of star formation within \citep[see][]{rapid-formation-mcs-hartmann-2001,mc-formation-stars-vazquez-semadeni-2007,
clumps-mhd-mc-formation-banerjee-2009,filaments-in-simulations-mc-formation-gomez-vazquez-semadeni-2014} appears plausible.

One example of such a large-scale flow is an expanding shell or super-shell around e.g. an OB-association or in general
driven by multiple stellar feedback of a star cluster or association \citep[see review by][]{supershells-mc-formation-review-dawson-2013}.
Molecular clouds may then either form inside the wall of such a shell \citep{mcs-in-supershells-dawson-2011} or at the interface
region when two such shells collide with each other. The latter has been recently investigated by \cite{gmc-formation-colliding-shells-dawson-2015}
for a young giant molecular cloud (GMC) at the interface of two colliding super-shells. From the comparison
of CO observations with high-resolution 3D hydrodynamical simulations they found that the GMC assembled into its current
form by the action of the shells.

The Scorpius-Centaurus OB-association \citep[Sco-Cen]{sco-cen-blaauw-1964,sco-cen-hipparcos-dezeeuw-1999,usco-full-population-preibisch-2002,sco-cen-sfhb-preibisch-mamajek-2008}
is the closest site of recent massive star formation to us and it consists of three sub-groups with different ages and well known stellar populations down to $2\,M_\odot$
\citep{sco-cen-stars-hipparcos-debruijne-1999}. The oldest one is the the Upper Centaurus-Lupus (UCL) sub-group with an age of $\sim 17$~Myr
harboring 66 B-stars. With an age of $\sim 15$~Myr the Lower Centaurus Crux (LCC) sub-group is somewhat younger and contains 42 B-stars.
The youngest sub-group is Upper-Scorpius (USco) with an age of $\sim 5$~Myr and consisting of 49 B-stars.

The feedback of the numerous massive stars in Sco-Cen probably
cleared the inner region of the association from diffuse matter creating
expanding loop-like  H\,{\rm I} shells around each of the sub-groups of the association 
\citep{sco-cen-HI-degeus-1992}. 
At the edge of the USco shell several dense molecular clouds with very young ($\leq 1-2$~Myr)
stellar populations are found. Of those the most prominent ones are the Lupus~{\rm I} cloud
(near the western edge of the shell) and the $\rho$~Oph cloud (near the eastern edge).

The Lupus~{\rm I} molecular cloud complex \citep[for an overview see][]{lupus-clouds-sfhb}
is found at a distance of 150~pc and consists of a $\approx2.6\degr\times0.6\degr$ (corresponds to
a physical size of $\approx6.8\times1.6$~pc) main filament extending in a north to south direction (Galactic coordinates)
and a ring-like structure of $\sim0.6\degr$ in diameter west of the main filament \citep[towards UCL; see][]{lupus-clouds-13co(2-1)-Tothill-2009}.
Recently \cite{lupus-submm-polarimetry-matthews-2014} noted also two smaller secondary filaments of 
which one is about half a degree long and runs perpendicular to the main
filament and seems to connect with it in the south. The other one is about a degree long and
lies south-west of the main filament extending from the southern end of the main filament to the ring-like structure.

In this work we concentrate our analysis only on the main filament commonly seen in 
all observations. We will refer to it as Lupus~{\rm I} or the Lupus~{\rm I} filament.

The cloud is found between the USco and the UCL sub-groups at a location where
the expanding USco shell appears to interact with a bubble currently powered by the winds of the remaining
B-stars of UCL. With its close distance Lupus~{\rm I} represents a good candidate where we can study how such
a collision process may form and influence new dense clouds in the ISM.

Lupus~{\rm I} has been mapped as part of several large surveys like the {\it Herschel}
Gould-belt survey \citep{herschel-gould-belt-andree-2010,lupus-clouds-herschel-rygl-2013} 
and the {\it Spitzer} Legacy Program 'From molecular clouds to planet-forming disks' 
\citep[c2d;][]{lupus-clouds-spitzer-c2d-chapman-2007,lupus-clouds-spitzer-c2d-merin-2008}.
These near-infrared to far-infrared surveys revealed the population of young stellar objects (YSOs)
within the cloud showing that it is dominated by pre-stellar and protostellar cores
indicating an on-going star formation event. \cite{lupus-clouds-herschel-rygl-2013}
found that the star formation rate (SFR) is increasing over the past $0.5-1.5$~Myr and 
\cite{lupus-clouds-spitzer-c2d-merin-2008} estimated a SFR of $\rm 4.3\,M_\odot\,Myr^{-1}$
for Lupus~{\rm I} from their {\it Spitzer} data.

Extinction maps of Lupus~{\rm I} have been created by various authors using different methods.
\cite{lupus-extinction-star-counts-cambresy-1999} created an extinction map based on optical star counts with
a resolution of a few arc-minutes.
Using 2MASS data \cite{lupus-clouds-spitzer-c2d-chapman-2007} created a visual extinction map with
a $2\arcmin$ resolution.
Also from 2MASS data a wide field extinction map of Lupus~{\rm I} has been presented by \cite{lupus-2mass-extinction-maps-lombardi-2008}.
It allowed extinction measurements down to $A_K=0.05$~mag, but had a resolution of $3\arcmin$.
\cite{lupus-clouds-spitzer-c2d-merin-2008} created extinction maps from their {\it Spitzer}
data by estimating the visual extinction towards each source classified as a background
star, based on their spectral energy distribution (SED) from 1.25 to $24\,\mu$m. Their maps
had a resolution of $2-5\arcmin$.
\cite{lupus-clouds-herschel-rygl-2013} created a column density map of dust emission in Lupus~{\rm I}
from their {\it Herschel} data by making a modified black body fit for each pixel in the
four bands from $160-500\,\mu$m. The resulting resolution was $36\arcsec$.

Here we present a far-infrared and sub-mm analysis of the Lupus~{\rm I} cloud based on newly obtained APEX/LABOCA sub-mm continuum data
at $870\,\mu$m and complementing those with {\it Herschel} and {\it Planck} archival data. In Sec.~\ref{sec:observations}
we briefly describe the observations used and their data reduction process. Sec.~\ref{sec:methods} gives an overview
of the methods used to create column density maps and probability distribution functions (PDFs) from the observations.
We present and discuss our results in Sec.~\ref{sec:results}. In Sec.~\ref{sec:surrounding} we consider the surroundings of Lupus~{\rm I} and the influence
of large-scale processes on the cloud. Finally, we summarize and conclude our paper in Sec.~\ref{sec:summary}.


\section{Observations and data reduction}
\label{sec:observations}

\subsection{LABOCA data}

The sub-mm continuum observations of Lupus~{\rm I} were performed with the APEX 12-m telescope located in the Chilean Atacama 
desert \citep{apex}. We used the LArge APEX BOlometer CAmera \citep[LABOCA,][]{laboca} which operates 
in the atmospheric window at $870\,\mu$m (345~GHz).
The angular resolution is 19.2\arcsec (HPBW), and the total field of view is 11.4\arcmin. 
At the distance of Lupus~{\rm I} (150~pc) the angular resolution corresponds to a spatial 
scale of $\sim2800$\,AU ($\sim0.01$\,pc). This is sufficient
to resolve the structure of molecular cores.

The LABOCA observations of Lupus~{\rm I} were obtained in Max-Planck and ESO Periods 91 and 92, on 24th March, 
31st August, 12th, 13th September 2013, and 31st March 2014 (PI: B.~Gaczkowski).
They were performed in the on-the-fly (OTF) mode,
scanning perpendicular as well as along the filament's major axis with a 
random position angle to the axis for each scan to reduce striping effects and improve
the sampling. The total observing time was 11.3 hours and the weather conditions good to average.

Data reduction included standard steps for sub-millimeter bolometer data,
using the {\tt BoA} software package \citep{boa}. First, data were converted from instrumental
count values to a Jansky scale using a standard conversion factor, then a
flatfield correction (derived from scans of bright, compact sources) was
applied. Corrections for atmospheric opacity were derived from skydips taken every 1-2 hours, and
finally residual correction factors were determined from observations of
planets and secondary calibrator sources. Data at the turning points of the map
were flagged, as well as spikes. The data from the individual 
bolometers were then corrected for slow amplitude drifts due to instrumental
effects and atmosphere by subtracting low-order polynomials from the 
time-stream data. Short timescale sky brightness variations (sky-noise)
was corrected by removing the correlated (over a large number of bolometers)
signal in an iterative fashion. Then the time-stream data were converted into
sky-brightness maps for each scan, and finally all scans combined into one
map.

\begin{figure}[htb]
 \centering
 \includegraphics[width=0.4\textwidth, keepaspectratio]{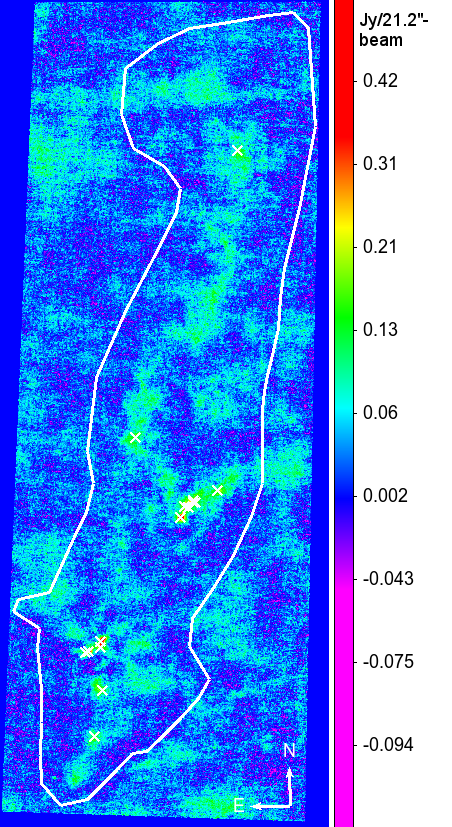}
 \caption{Lupus~{\rm I} LABOCA $870\,\mu$m map with square-root intensity scaling in units of Jy/$21.2\arcsec$-beam. The positions of the 15 cores that were found by {\tt Clumpfind} are marked with white crosses. The white polygon marks the common area within which the mass of the cloud was calculated from the LABOCA and {\it Herschel} column density maps (see Section~\ref{sec:total-mass-lupus}).}
\label{img:laboca-map}
\end{figure}

The removal of the correlated sky variations is known to filter out also
astronomical emission from extended sources, usually leading to negative
artifacts surrounding bright emission structures. In order to recover some
of the extended emission, an iterative source modeling procedure was applied,
using the result from the previous iteration to construct an input model for
the following iteration. To construct the model, the previous map was smoothed,
all pixels below a pixel value of 0 set to 0, and the map smoothed
again. In the following iteration, the model is subtracted from the time-stream
data before de-spiking, baseline subtraction, and sky-noise removal, and added
back to the data stream afterwards, before the new map is created. This
procedure effectively injects artificial flux into the mapped area
and is prone to create runaway high surface brightness areas especially in
regions without significant astronomical signal and low signal-to-noise.
For that reason, the model image was set entirely to 0 in areas of the map
with poor coverage, below a certain threshold in rms (basically the map edges),
and the procedure was stopped after 25 iterations. Typically, structures on
scales of the order of $3-4\arcmin$ can be recovered, larger scales get more
and more filtered out \citep[see simulations done by][]{chamaeleon-laboca-belloche-2011}.
The gridding was done with a cell size of $6.1\arcsec$ and the
map was smoothed with a Gaussian kernel of size $9\arcsec$ (FWHM).
The angular resolution of the final map is $21.2\arcsec$ (HPBW) and the
rms noise level is 23~mJy/$21.2\arcsec$-beam (hereafter the notation per 
beam in context of LABOCA will mean per $21.2\arcsec$-beam). The resulting
map is shown in Fig.~\ref{img:laboca-map}.

In order to determine the total sub-mm flux of the entire LABOCA map, we integrated
all pixel values above the $3\sigma$ noise level of 69~mJy/beam.
This yielded a total flux value of 476~Jy. Our peak intensity in the map 
is $I^\mathrm{max}_\nu=1.37$~Jy/beam.

\subsection{Herschel archival data}

The Lupus~{\rm I} cloud complex was observed by the {\it Herschel} far-infrared observatory \citep{herschel} in January 2011
 as part of the Gould-belt survey \citep[][]{herschel-gould-belt-andree-2010,lupus-clouds-herschel-rygl-2013}.
The Photodetector Array Camera and Spectrometer \citep[PACS,][]{pacs} and the Spectral and Photometric Imaging
Receiver \citep[SPIRE,][]{spire} were used to map an area of $\approx2\degr \times 2.3\degr$.
We retrieved these data from the {\it Herschel} science archive and reduced them
using HIPE v12.1 \citep{hipe} for the calibration and the deglitching of the Level~0 PACS and SPIRE data.
The Level~1 data of both instruments were then used to produce the final maps with the 
Scanamorphos package v24 \citep[][/parallel and /galactic options were switched on in both cases
to preserve extended structures in complex, bright Galactic fields]{scanamorphos}. 
The pixel-sizes for the five maps at 70, 160, 250, 350 and 500\,$\mu$m 
were chosen as $3.2\arcsec$, $4.5\arcsec$, $6\arcsec$, $8\arcsec$, and $11.5\arcsec$, respectively.

The absolute calibration of the {\it Herschel} data was done following the approach described in \cite{dust-temperature-herschel-planck-bernard-2010}.
Using the {\it Planck} and the IRAS data of the same field as the {\it Herschel} data
one calculates the expected fluxes to be observed by {\it Herschel} at each band and from
that computes the zero-level offsets (see Table~\ref{tbl:herschel-offsets}). Those are then added
to the {\it Herschel} maps to create the final mosaics.

\begin{table}[htb]
 \centering
 \caption[]{Zero-level offsets for the {\it Herschel} maps obtained from the cross-correlation with Planck and IRAS.}
 \begin{tabular}{r c}
 \hline\hline
 \noalign{\smallskip}
  Band &  Offset \\  $[\,\mu{\rm m}]$  &  MJy $\rm sr^{-1}$ \\
 \noalign{\smallskip}
 \hline
 \noalign{\smallskip}
   70 &   5.68954 \\    
  160 &   14.3011 \\    
  250 &   8.11490  \\    
  350 &   4.06405  \\    
  500 &  1.68754   \\   
 \noalign{\smallskip}
 \hline
 \end{tabular}
\label{tbl:herschel-offsets}
\end{table}

For the dust properties and temperatures considered in our analysis the color corrections for both PACS and SPIRE
are of order 1-2\% and hence negligible.

\subsection{Planck archival data}

We retrieved the all-sky maps of the High-Frequency-Receiver (HFI)
in the 353, 545, and 857~GHz bands (corresponding to the FIR and sub-mm wavelengths
of 850, 550, and $350\,\mu$m) from the {\it Planck} legacy archive (release PR1 21.03.2013).
From these data cubes in HEALPix\footnote{\url{http://healpix.jpl.nasa.gov/html/intro.htm}} format we extracted for
each wavelength separately the area covering the Lupus~{\rm I} region and created gnomonic projected
maps. The resolution in each of the three bands is $5\arcmin$ and the pixel size was chosen as $1.7\arcmin$.

\section{Data analysis}
\label{sec:methods}

\subsection{Column density and temperature maps of Lupus~{\rm I}}

From our three data sets of LABOCA, {\it Herschel}, and {\it Planck} we calculated column density maps
to characterize the cloud in a multi-wavelength approach. Additionally, we constructed a temperature map
from the {\it Herschel} data.

\subsubsection{Column density and temperature map from {\it Herschel} SED fit with all SPIRE bands}

The standard and often practiced way to derive column density and temperature maps from the {\it Herschel} data
is to fit an SED to the observed fluxes of the {\it Herschel} bands for each pixel of the maps \citep[see e.g.][]{herschel-chamaeleon-sed-fit-oliveira-2014,herschel-planck-column-density-maps-lombardi-2014,herschel-physical-dust-properties-massive-sf-regions-battersby-2014}.
Here we fit the SED to the three SPIRE bands 250, 350, and $500\,\mu$m. We do not include the two PACS 70 and $160\,\mu$m bands, because
those observations were corrupted by stray moonlight and hence are not reliable for an analysis of large-scale structures.

Because Lupus~{\rm I} is optically thin to the dust emission at the considered densities
and wavelengths 
we can model its emission as a modified black body. We assume that the long wavelength emission
($\lambda\ge 250\,\mu$m) of a pixel [i,\,j] comes from a unique
species of grains being all at the same equilibrium temperature, and having a power-law wavelength
dependent opacity.
If $L_{\nu}^{i,j}(\lambda)$ is the monochromatic luminosity of pixel [i,\,j] at wavelength $\lambda$,
then it can be expressed as

\begin{equation}
L^{i,j}_{\nu}(\lambda) = M^{i,j}_\mathrm{d}\times\kappa(\lambda_0)\left(\frac{\lambda_{0}}{\lambda}\right)^\beta\times4\pi\,B_\nu(\lambda,T^{i,j}_\mathrm{d})
\end{equation}
where $\kappa$ is the dust opacity, $\beta$ the emissivity
index, $B_\nu(\lambda,T_\mathrm{d})$ the black body spectral flux density
for a dust temperature $T_\mathrm{d}$. 
The two free parameters $M^{i,j}_\mathrm{d}$ and $T^{i,j}_\mathrm{d}$ are the dust mass and temperature per pixel of the
material along the line-of-sight.
Here we adopt a typical $\kappa(\lambda_0)=5.91\,{\rm cm^2\,g^{-1}}$ with $\lambda_0 = 350\,\mu$m for dust grains with thin
ice mantels and gas densities $<10^5$\,cm$^{-3}$ \citep{dust-opacities-for-ps-cores-ossenkopf-1994}.
Using the available information about $\beta$ from the {\it Planck} data, we found that the emissivity index
within the Lupus~{\rm I} cloud lies between $\sim1.6$ and $\sim1.7$. Therefore we fixed its value to $\beta=1.65$ as an average value
within the cloud. Considering the low resolution of the $\beta$ map of $30\arcmin$ this is a reasonable approximation.
In this way we also limit the number of free parameters in the fit making it more stable.

\begin{figure*}[htb]
 \subfigure{
  \centering
   \includegraphics[height=0.5\textwidth, keepaspectratio]{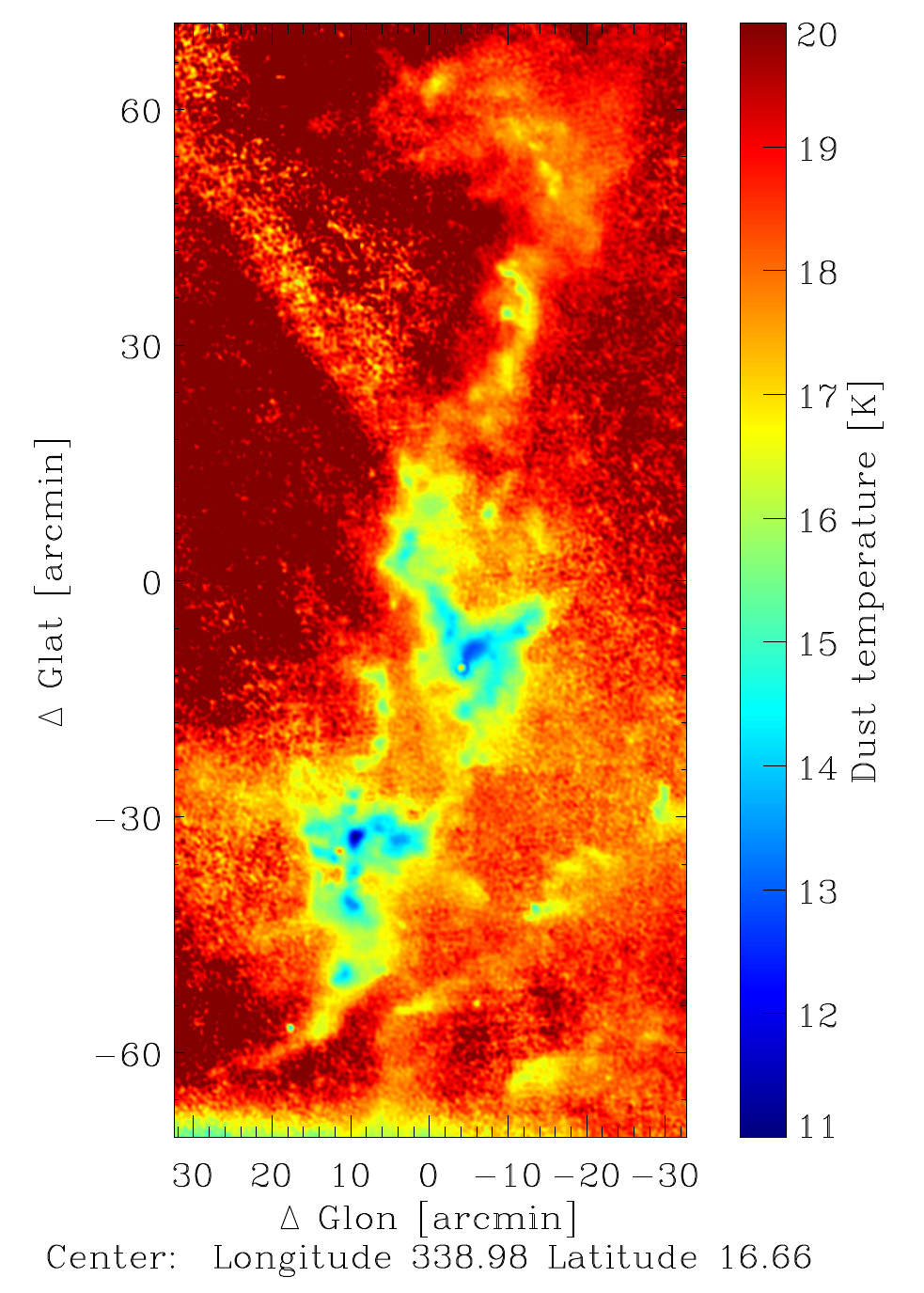}
    \label{img:herschel-temp-map}}
 \subfigure{
  \centering
  \includegraphics[height=0.38\textheight, keepaspectratio]{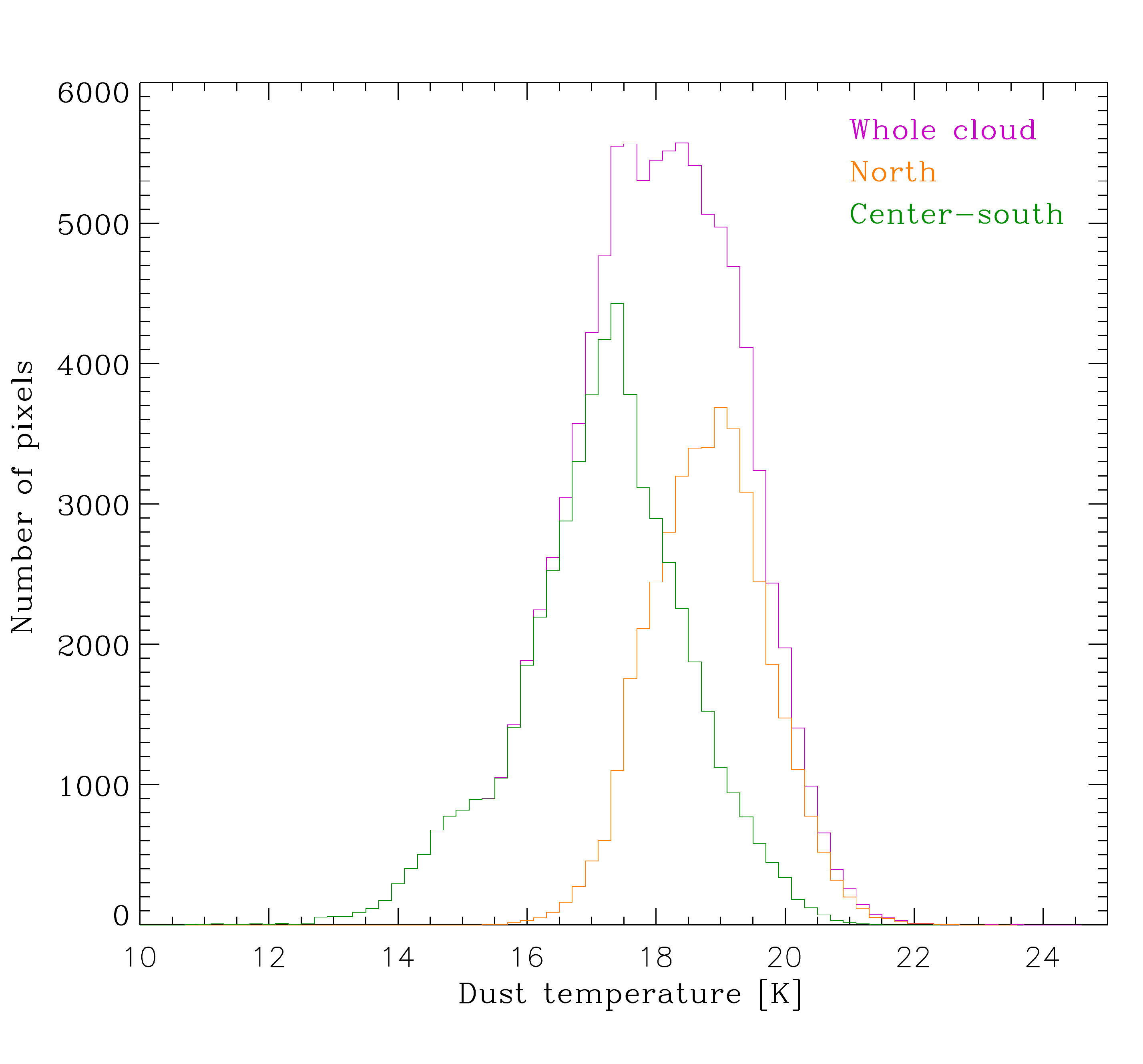}
  \label{img:herschel-temp-histo}}
 \caption{{\it Left:} Temperature map of Lupus~{\rm I} from the {\it Herschel} SPIRE SED fit. {\it Right:} Histogram of the dust temperature for 
the whole cloud (purple),
the northern (orange), and the center-south part (green).}
\end{figure*}

After convolving the 250 and $350\,\mu$m maps to the resolution of the $500\,\mu$m
band using the kernels from \cite{convolution-kernels-aniano-2011}, the modified black
body fit was performed pixel-by-pixel. From the dust mass $M^{i,j}_\mathrm{d}$ in each pixel the total column density 
for both dust and gas was then calculated as

\begin{equation}
N^\mathrm{H_{SED}}_\mathrm{H_2} = \frac{M_{i,j}\,R}{\mu_\mathrm{H_2}\,m_H}
\end{equation}
where $R = 100$ is the gas-to-dust mass ratio, $\mu_\mathrm{H_2}=2.8$ the molecular weight per hydrogen molecule, 
and $m_\mathrm{H}$ the hydrogen atom mass. 
The resulting dust temperature values $T^{i,j}_\mathrm{d}$ from the fit at each pixel give the temperature map
of Lupus~{\rm I} (shown in Fig.~\ref{img:herschel-temp-map}) at the resolution of the $500\,\mu$m band, i.e. $\rm FWHM_{500}=36\arcsec$.

Since the composition of the dust grains and their density is unknown,
the choice of a particular dust model and thus a specific opacity law introduces an
uncertainty of the dust opacity and thus of the resulting column density.
In order to estimate it, we took different $\kappa(\lambda_0)$ values at $350\,\mu$m for grains with and without ice mantles
and the three initial gas densities of $<10^5$, $10^5$, and $10^6$\,cm$^{-3}$ in the online table of
\cite{dust-opacities-for-ps-cores-ossenkopf-1994}. They vary between $3.64\,{\rm cm^2\,g^{-1}}$ (MRN\footnote{Mathis-Rumpl-Nordsieck size
distribution of interstellar grains \citep{dust-opacity-mrn-1977}}
without ice mantles and $\rho<10^5$\,cm$^{-3}$) and $11.3\,{\rm cm^2\,g^{-1}}$ (MRN with
thin ice mantles and $\rho=10^6$\,cm$^{-3}$). Hence, our chosen value of $\kappa(\lambda_0)$ might still vary
by a factor of about 2.

The statistical error on both the final column density map and the temperature map consists of errors of the calibration, the photometry,
and the SED fitting process. We conservatively estimated the sum of these uncertainties to be $\sim20\%$ for both maps.

\subsubsection{Column density map from {\it Herschel} SPIRE $250\,\mu$m map}

To obtain a column density map with the resolution of the SPIRE $250\,\mu$m band (i.e. $\rm FWHM_{250}=18\arcsec$),
we followed the technique described in \cite{column-density-maps-herschel-juvela-2013}. Using the intensity map of the SPIRE $250\,\mu$m band
and the previously calculated temperature map (at the lower resolution) to compute the Planck function $B_\nu(T_\mathrm{d})$, it is possible to gain another factor of two in resolution
compared to the SED fitting (to all three SPIRE bands) case. In this way the column density simply is

\begin{equation}
N^\mathrm{H_{250}}_\mathrm{H_2} = \frac{I_{250\,\mu{\rm m}}^{i,j}\,R}{B_\nu(250\,\mu{\rm m},T_\mathrm{d}^{i,j})\,\kappa_{250\,\mu{\rm m}}\,\mu_\mathrm{H_2}\,m_H}
\label{eq:NH2}
\end{equation}
with $I_{250\,\mu{\rm m}}^{i,j}$ the intensity of the SPIRE $250\,\mu$m band.
We used this higher resolution {\it Herschel} column density map for our further analysis of Lupus~{\rm I}.

\subsubsection{Column density map from the LABOCA data}
\label{sec:col-dens-map-Laboca}

Lupus~{\rm I} is sufficiently far away from ionizing UV sources to exclude significant amounts of free-free emission.
We may therefore assume the sub-mm fluxes to be entirely due to thermal dust emission.
To determine whether the cloud is optically thin in this wavelength regime
one can look at the peak intensity in the LABOCA map, which is $I^{\rm max}_\nu=1.37$~Jy/beam.
The corresponding optical depth can be calculated via formula (3) of \cite{atlasgal}.

\begin{equation}
\tau_{870\,\mu\mathrm{m}}=-\ln\left[ 1-\frac{I_\nu}{\Omega\,B_\nu(T_\mathrm{d})}\right]
\end{equation}
where $\Omega$ is the beam solid angle.
In the case of Lupus~{\rm I} this yields values of $\tau_{870\,\mu{\rm m}}\le0.01$ for temperatures $T_\mathrm{d}\ge10$~K. Hence
the cloud is clearly optically thin in this wavelength regime.
Therefore, the $870\,\mu$m intensities are directly proportional to the column densities of
the interstellar dust and the line-of-sight extinction. They can be converted to the
beam-averaged hydrogen molecule column density via equation~(\ref{eq:NH2}).
 
We again assume a gas-to-dust mass ratio of $R = 100$ and for consistency extrapolate our dust
model used for the {\it Herschel} analysis to the LABOCA wavelength, which yields a value of 
$\kappa_{870\,\mu{\rm m}} = 1.32\,{\rm cm}^2\,{\rm g}^{-1}$. Because the assumption of a constant dust temperature
throughout the cloud is not valid, we used the temperature map from the {\it Herschel} analysis to calculate
the black body function at each pixel of the map. In this way one obtains a column density map at
the original resolution of the LABOCA map (i.e. $\rm FWHM_{870}=21.2\arcsec$).

\subsubsection{Column density and temperature map from the {\it Planck} data}

The thermal emission of interstellar dust over the whole sky was captured by the HFI-instrument of {\it Planck} 
at its 6 available wavelengths between $350\,\mu$m and 3\,mm. Together with the IRAS $100\,\mu$m data the emission
can be well modeled by a modified black body. The details of the model and the fitting procedure are described in
\cite{planck-2013-sky-thermal-dust-emission-model}. The resulting maps of temperature and optical depth (at $850\,\mu$m)
can be downloaded from the {\it Planck} archive\footnote{\url{http://irsa.ipac.caltech.edu/data/Planck/release_1/all-sky-maps/maps/HFI_CompMap_ThermalDustModel_2048_R1.20.fits}}. We extracted the part covering Lupus~{\rm I} from both maps
to get gnomonic projected maps. The column density map was then computed from the optical depth map using
the relation 

\begin{equation} 
N_\mathrm{H_2} = \frac{5.8\times10^{21}\,{\rm cm^{-2}}}{2}\times0.497\times10^4\,\tau_{\rm 353\,GHz}
\end{equation}
suggested in \cite{planck-2013-sky-thermal-dust-emission-model} for regions with higher column density than the diffuse ISM,
i.e. molecular clouds.

\subsection{Column density PDFs}

Probability distribution functions (PDFs hereafter) of the column density
are a widely used way to characterize the evolution and state of molecular clouds.
Both simulators \citep[see e.g.][]{pdf-simulations-turbulent-clouds-federrath-klessen-2012,
pdf-column-density-simu-ward-2014,pdf-column-density-evolution-numerical-girichidis-2014} and 
observers \citep[see e.g.][]{pdf-column-density-kainulainen-2009,pdf-column-density-kainulainen-2011,
herschel-column-density-pdf-taurus-palmeirim-2013,herschel-hii-compression-pdf-tremblin-2014,
pdf-column-density-schneider-2015,pdf-serpens-core-roccatagliata-2015}
use them in their studies and they are
a good tool to compare both simulations and observations with each other.

As Lupus~{\rm I} is at near distance to us and lies at a high Galactic latitude, 
far away from the Galactic plane, we expect the contamination of the map by overlaying
foreground or background emission to be small.
We normalize the PDFs to the dimensionless $\eta=\ln( N_\mathrm{H_2}/\langle N_\mathrm{H_2}\rangle)$
which gives the opportunity to compare different regions of different column density, as well as 
the same cloud, but observed with different instruments. 
For similar observations of the Orion clouds \citet{pdf-column-density-orion-schneider-2013} showed
that the effect of varying the resolution of the maps within $18\arcsec$ and $36\arcsec$ did not affect
the main features of the PDFs (shape, width, etc.) significantly. The Lupus~{\rm I} cloud is much closer,
yielding a much finer physical resolution and the tests we performed on our maps smoothing them with a grid of Gaussians with different
FWHM and rebinning them to several different pixel sizes confirmed their findings. If the difference,
however, is as big as e.g. the one in angular resolution between {\it Herschel} and {\it Planck},
it is not longer possible to maintain the features of the PDF unchanged when degrading the {\it Herschel} map
by a factor of $\approx8$ (from $500\,\mu$m) to $\approx17$ (from $250\,\mu$m). But in our case both the
angular resolution and the pixel size of the {\it Herschel} and the LABOCA column density map are
almost the same. Therefore, both maps have sufficiently large numbers of pixels to have a high statistic.

The basic shape of the PDF in the low column density regime 
can be described by a lognormal distribution.
In more complicated cases
the PDF might also be the combination of two lognormals:

\begin{equation}\label{eq_fit}
p(\eta)=\frac{\epsilon_1}{\sqrt{2\pi\sigma_1^2}}\exp\left(\frac{-(\eta-\mu_1)^2}{2\sigma_1^2}\right)+\frac{\epsilon_2}{\sqrt{2\pi\sigma_2^2}}\exp\left(\frac{-(\eta-\mu_2)^2}{2\sigma_2^2}\right)
\end{equation}
where $\epsilon_i$ is the norm, 
$\mu_i$ and $\sigma_i$ are the mean logarithmic column density and dispersion
of each lognormal. In the high column density regime one often finds a deviation
from the lognormal.
This can be modeled with a power-law of slope $s$ which is equivalent to the slope of a spherical density profile
$\rho(r) \propto r^{-\alpha}$ with $\alpha = -2/s + 1$ \citep[see][]{sf-efficiency-in-turbulent-magnetized-clouds-federrath-klessen-2013,pdf-column-density-orion-schneider-2013}.

Using a least-square method (Levenberg Marquardt algorithm; Poisson weighting)\footnote{{\tt MPCURVEFIT} routine for IDL}
we derived the characteristic values of the distributions by fitting the PDFs with either one lognormal 
to the distribution around the single peak and where applicable a power-law to the high density tail,
or with two lognormals to the distributions around the first and second peak, respectively, and a possible power-law tail.
We checked the robustness of the fits by performing it on four different binsizes for the histograms (0.05, 0.1, 0.15, 0.2)
and looking at the variation of the resulting fit parameters which was of the order $2\%-8\%$. We took
the final parameters from the fits for a binsize of 0.1 \citep[see][]{pdf-column-density-schneider-2015} and conservatively adopted an error of 10\% for all fit parameters. The results are summarized in Table~\ref{tbl:PDF-fit-parameters}.

\begin{table*}[htb]
 \centering
 \caption[]{Results of the fits to the {\it Herschel} PDFs of Fig.~\ref{img:PDFs}.}
 \begin{tabular}{c c c c c c c c c}
 \hline\hline
 \noalign{\smallskip}
  Region &  $\sigma_1$ &  $\rm N_{{H_2},peak1}$               &  $\sigma_2$  &  $\rm N_{{H_2},peak2}$              & $<\rm N_\mathrm{H_2}>$                   & $\rm N_{H_2,dev}$                   & s & $\alpha$ \\
         &             &  [$\times10^{21}\,{\rm cm^{-2}}$]    &              &  [$\times10^{21}\,{\rm cm^{-2}}$]   & [$\times10^{21}\,{\rm cm^{-2}}$]  & [$\times10^{21}\,{\rm cm^{-2}}$]  &   &          \\
 \noalign{\smallskip}
 \hline
 \noalign{\smallskip}
   Whole cloud  &   0.56 &  0.866 &   0.26  &  2.57  &  1.44  & 3.60  & -2.55 & 1.78 \\    
   Center-south &   0.46 &  0.934 &   0.35  &  2.71  &  1.72  & 4.50  & -2.43 & 1.82 \\  
   north        &   0.55 &  0.794 &   0.18  &  2.23  &  1.10  & --    & --    & --   \\    
 \noalign{\smallskip}
 \hline
 \end{tabular}
\label{tbl:PDF-fit-parameters}
\end{table*}

\section{Results}
\label{sec:results}

\subsection{Column density and temperature maps}
\label{sec:results:Column density and temperature maps}

\begin{figure*}[htb]
\centering
 \subfigure[]{
  \centering
  \includegraphics[height=0.4\textheight, keepaspectratio]{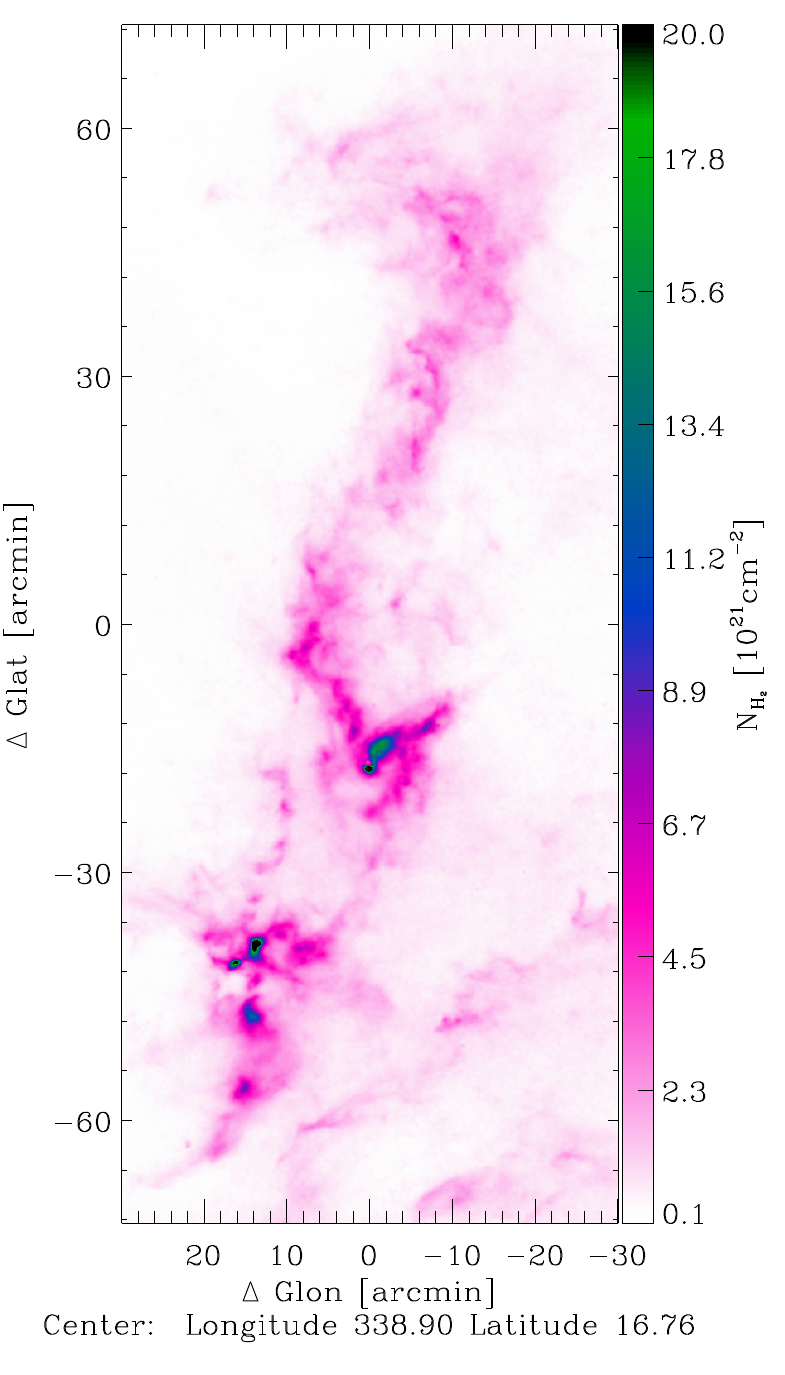}
  \label{img:Herschel_coldens_high}}
 \subfigure[]{
  \centering
  \includegraphics[height=0.4\textheight,keepaspectratio]{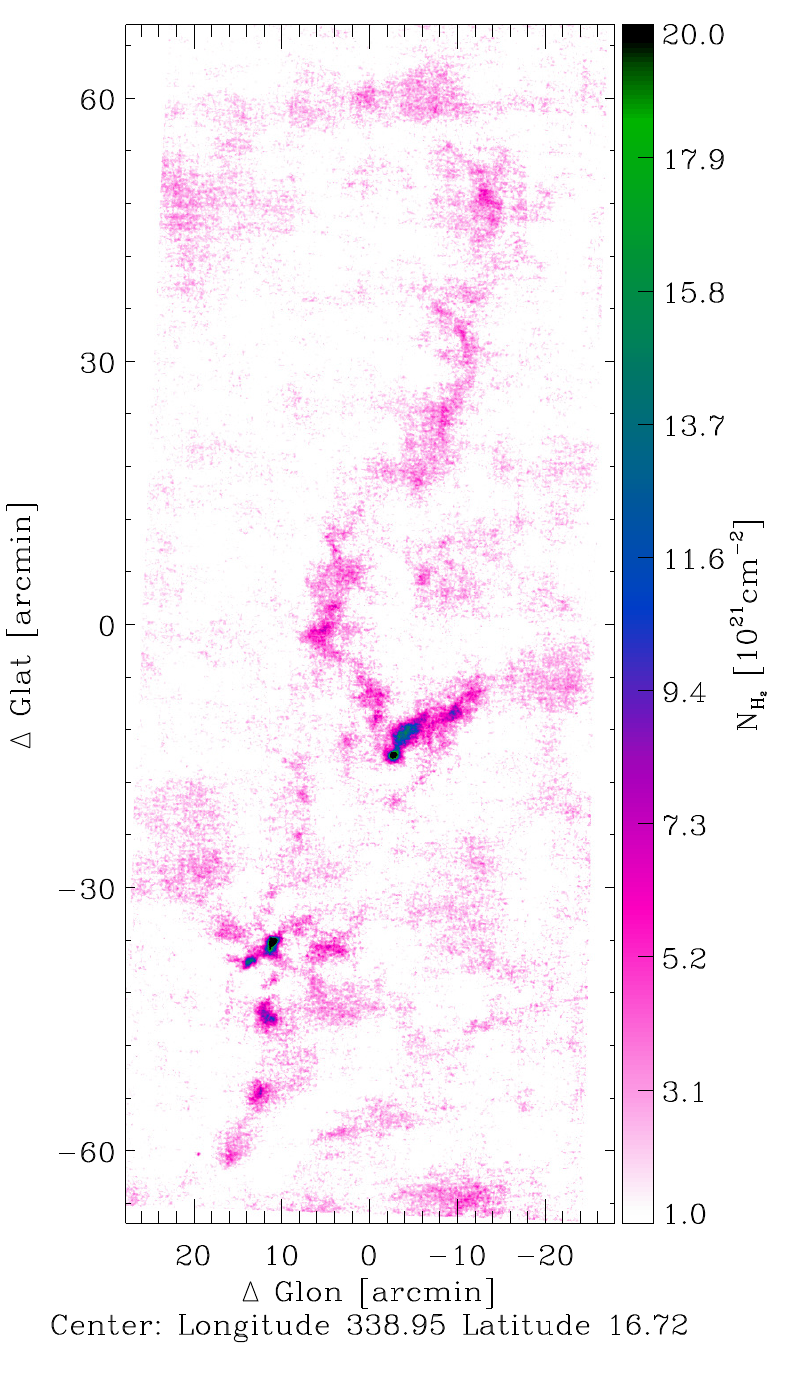}
  \label{img:Laboca_coldens}}
 \subfigure[]{
  \centering
  \includegraphics[height=0.4\textheight, keepaspectratio]{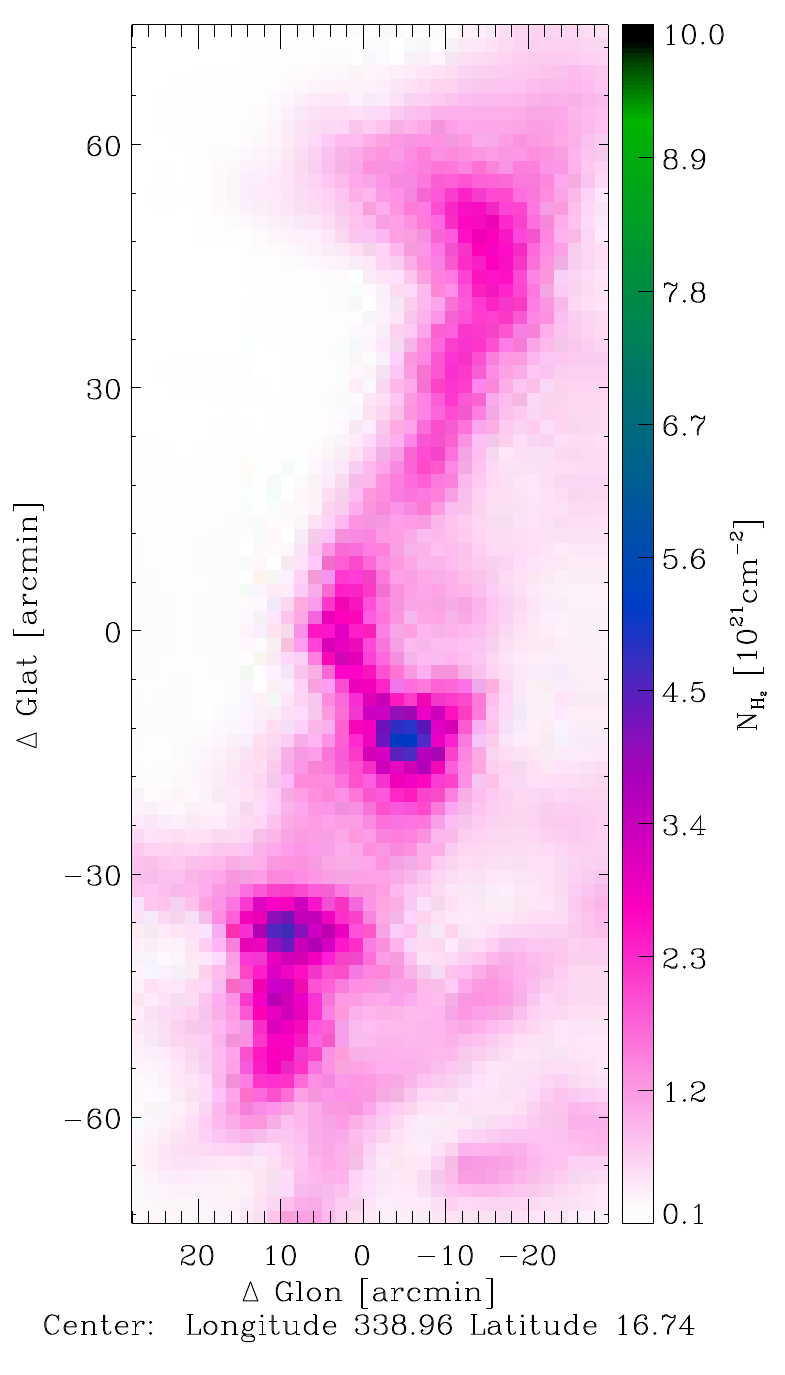}
  \label{img:Planck_coldens}}
\caption{Column density maps of Lupus~{\rm I} from (a) the {\it Herschel} $250\,\mu$m map, (b) LABOCA, and (c) {\it Planck}.}
\label{img:coldens-maps}
\end{figure*}

\begin{figure}[htb]
 \centering
   \includegraphics[width=0.5\textwidth, keepaspectratio]{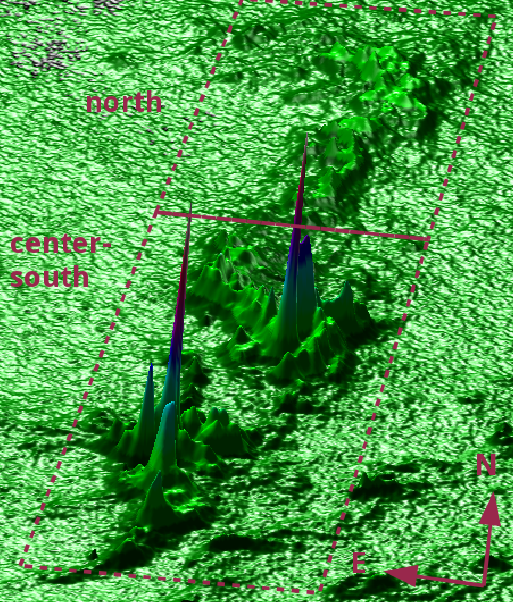}
 \caption[Column density PDFs of Lupus~{\rm I}.]{3D-surface plot of the {\it Herschel} $250\,\mu$m column density map. 
One can clearly identify the cores in the center-south of Lupus~{\rm I} and see the on average lower column density in the northern
part of the cloud. The red dashed boxes mark the two regions we distinguish in this work.}
\label{img:herschel-col-dens-map-3d}
\end{figure}

The {\it Herschel} column density map of Lupus~{\rm I} is shown in Fig.~\ref{img:Herschel_coldens_high}.
The average column density is <$\rm N_\mathrm{H_2}$> $\rm =1.44\times10^{21}\,cm^{-2}$.
The map reveals two distinct regions within the filament (see also Fig.~\ref{img:herschel-col-dens-map-3d}).
The north where the column densities are lowest 
(<$\rm N_\mathrm{H_2}$> $\rm =1.10\times10^{21}\,cm^{-2}$) and just one dense core can be seen \citep{lupus-clouds-herschel-rygl-2013} and the 
center-south part where several cores with high column densities
above $\rm 10^{22}\,cm^{-2}$ are found and the average column density (<$\rm N_\mathrm{H_2}$> $\rm =1.72\times10^{21}\,cm^{-2}$) 
is about 60\% higher than in the northern part.

Comparing our map to the one derived by \cite{lupus-clouds-herschel-rygl-2013} we note that our values are lower
by a factor of about $2-2.5$. This can be explained by the use of a different
dust model (opacity and dust spectral index) and the inclusion of the PACS $160\,\mu$m band in their work.
The extinction maps created from 2MASS and {\it Spitzer} data \citep[see][]{lupus-clouds-spitzer-c2d-chapman-2007,
lupus-2mass-extinction-maps-lombardi-2008,lupus-clouds-spitzer-c2d-merin-2008} have a resolution that is lower
by a factor of $\sim7-30$. Nevertheless, between those maps and our column density map is a clear resemblance.

The column density map of Lupus~{\rm I} we obtained from LABOCA (see Fig.~\ref{img:Laboca_coldens}) is dominated by the
dozen cores in the center-south part and the denser dust in the central part of the filament. The average column density
is <$\rm N_\mathrm{H_2}$> $\rm =1.47\times10^{21}\,cm^{-2}$.

The {\it Planck} column density map of Lupus~{\rm I} with a resolution of $5\arcmin$ reveals the basic structure
of the cloud which agrees well with the column density maps from LABOCA and {\it Herschel}. This is consistent with what
is shown by the other two maps when smoothed to the resolution of {\it Planck} and thus smoothing out the highest column density peaks.

The temperature map obtained from the {\it Herschel} SED fit (see Fig.~\ref{img:herschel-temp-map}) shows an anti-correlation of the temperature
with the column density. The densest parts are the coldest and the less dense the material the warmer
it becomes. Also one sees again a difference between the northern and the center-south part of the cloud.
In the north dust temperatures between $\approx15-23$~K with a mean and median of $\approx19$~K are found.
But only in the dense pre-stellar core and slightly
north of it the temperature drops down to $\approx15$~K.
The maximum temperature in the center-south part is $\approx22$~K at the edges of the filament.
The inner part of the filament has an average temperature of $\approx17$~K. But it drops
down to even 11~K in the densest cores.
From the histogram of the dust temperatures (see Fig.~\ref{img:herschel-temp-histo}) one can see that
most of the dust in the north (orange histogram) ranges between $\approx18-20$~K whereas in the center-south part (green) it
is between $\approx16-18$~K. This means that the dust in the center-south is on average 2~K colder than in the north.

\subsection{Column density PDFs}

\begin{figure*}[htb]
\vspace{-0.5cm}
\centering
\subfigure{
  \centering
  \includegraphics[height=0.28\textheight, keepaspectratio]{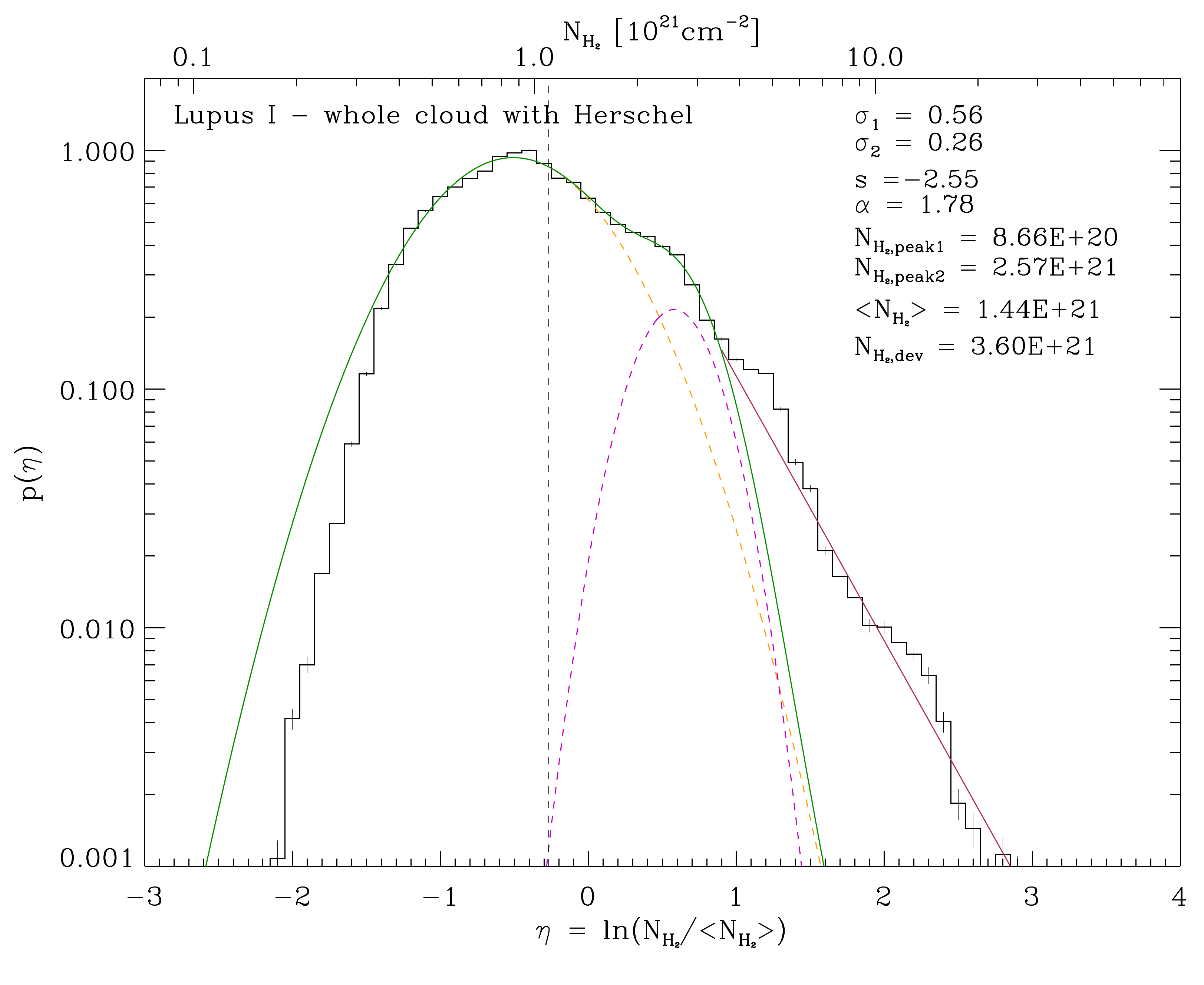}
  \label{img:Herschel_pdf_all}}
\subfigure{
  \centering
  \includegraphics[height=0.28\textheight, keepaspectratio]{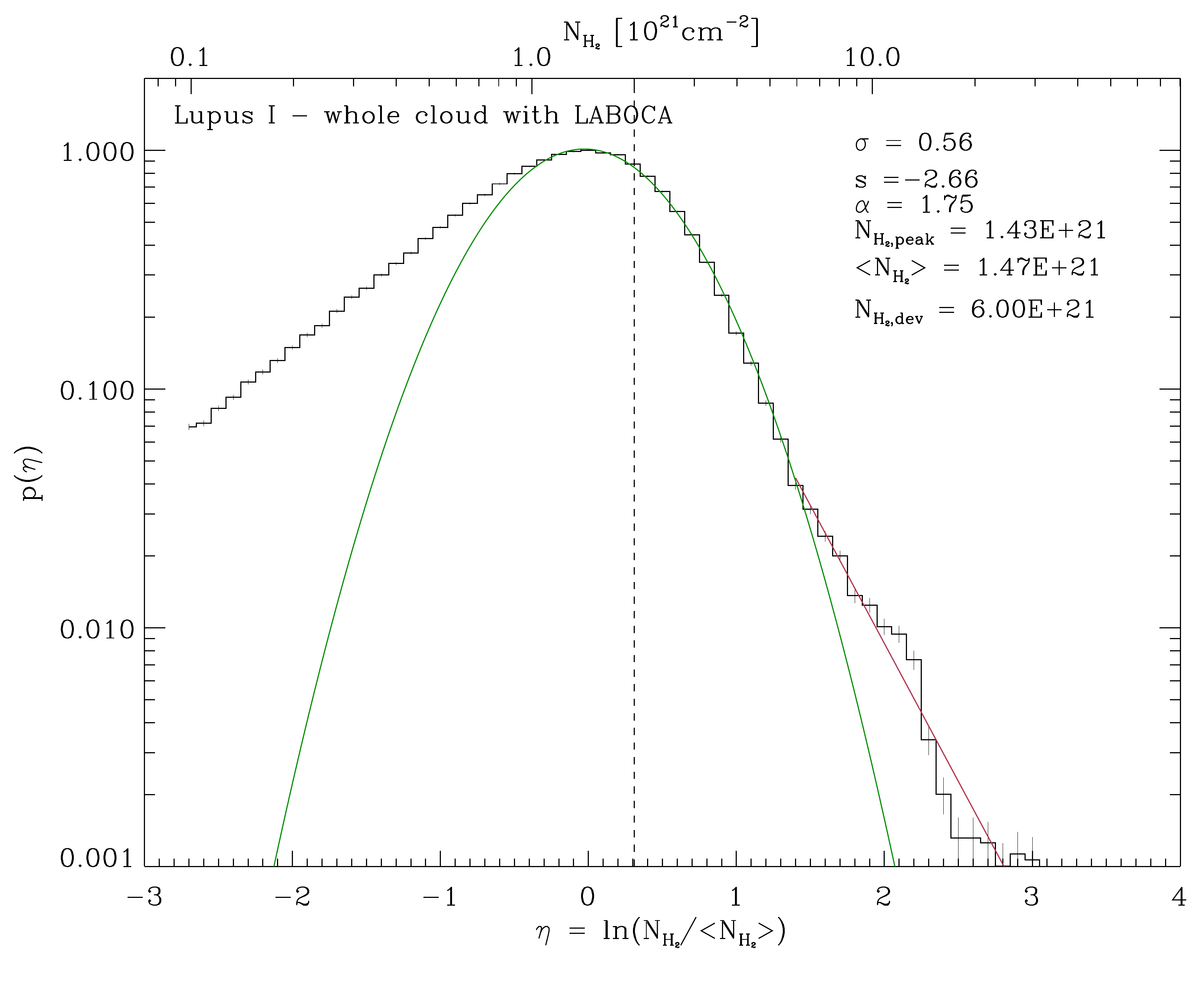}
  \label{img:Laboca_pdf_all}}
\subfigure{
  \centering
  \includegraphics[height=0.28\textheight,keepaspectratio]{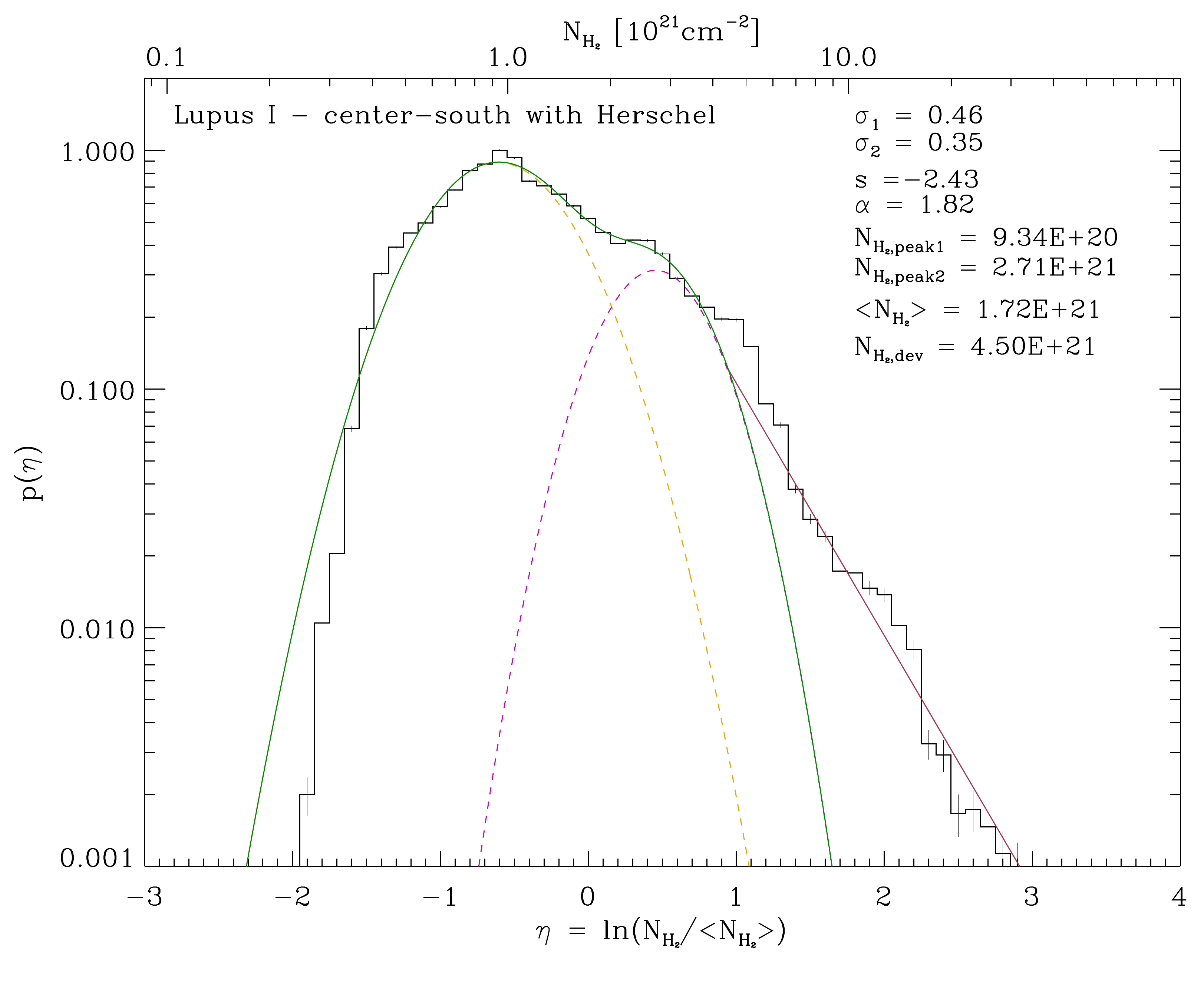}
  \label{img:Herschel_pdf_MS}}
\subfigure{
  \centering
  \includegraphics[height=0.28\textheight,keepaspectratio]{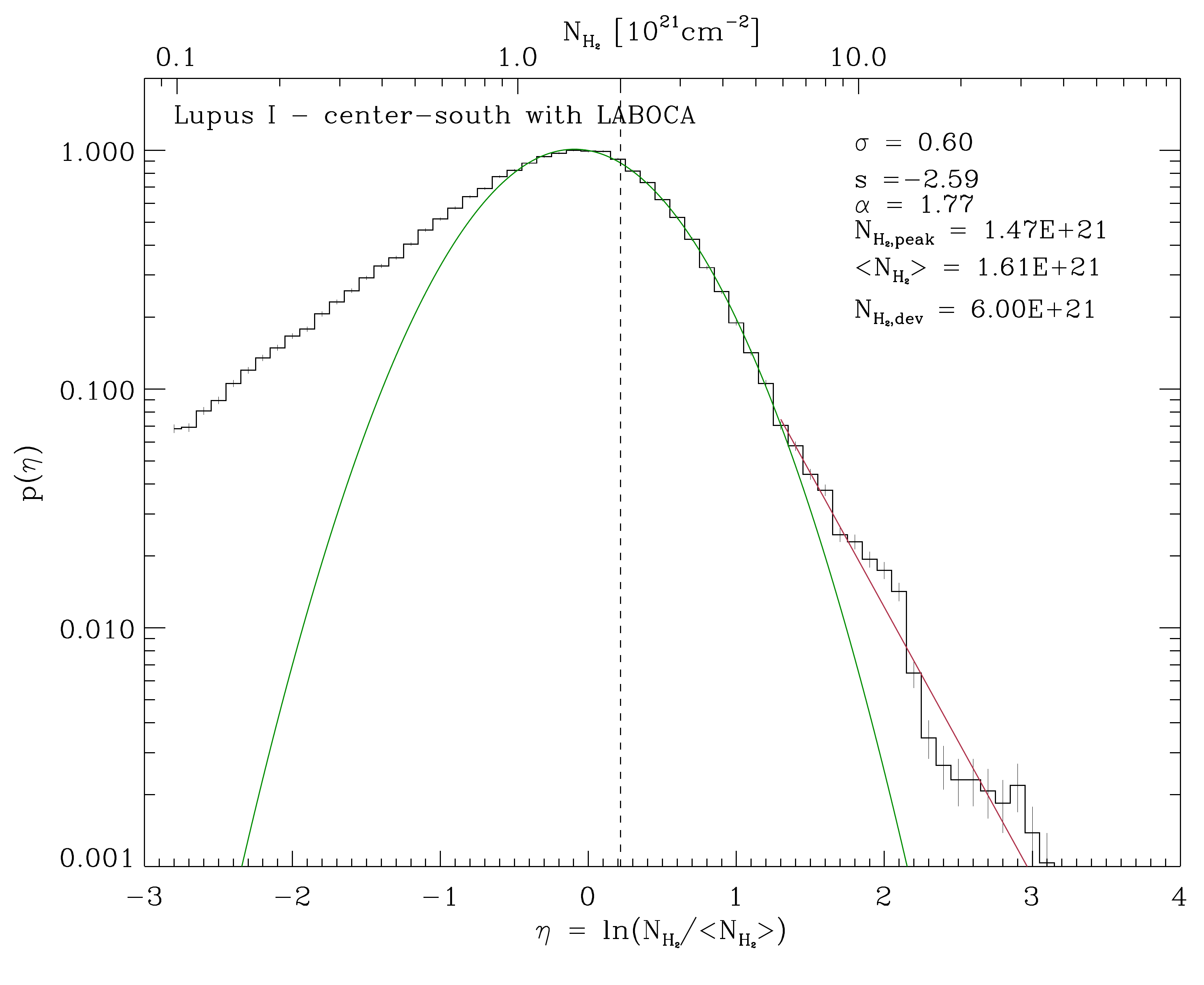}
  \label{img:Laboca_pdf_MS}}
\subfigure{
  \centering
  \includegraphics[height=0.28\textheight, keepaspectratio]{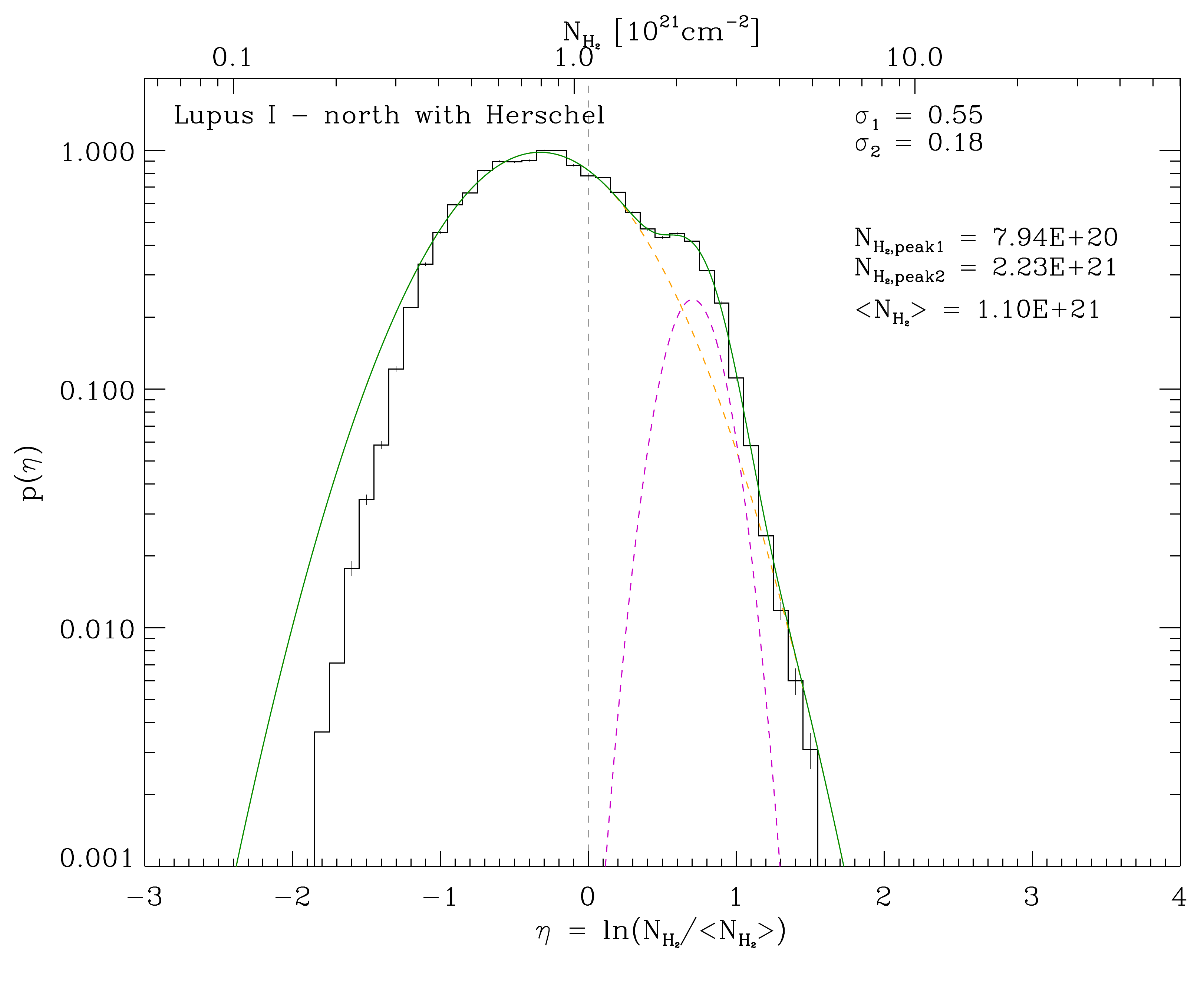}
  \label{img:Herschel_pdf_N}}
\subfigure{
  \centering
  \includegraphics[height=0.28\textheight, keepaspectratio]{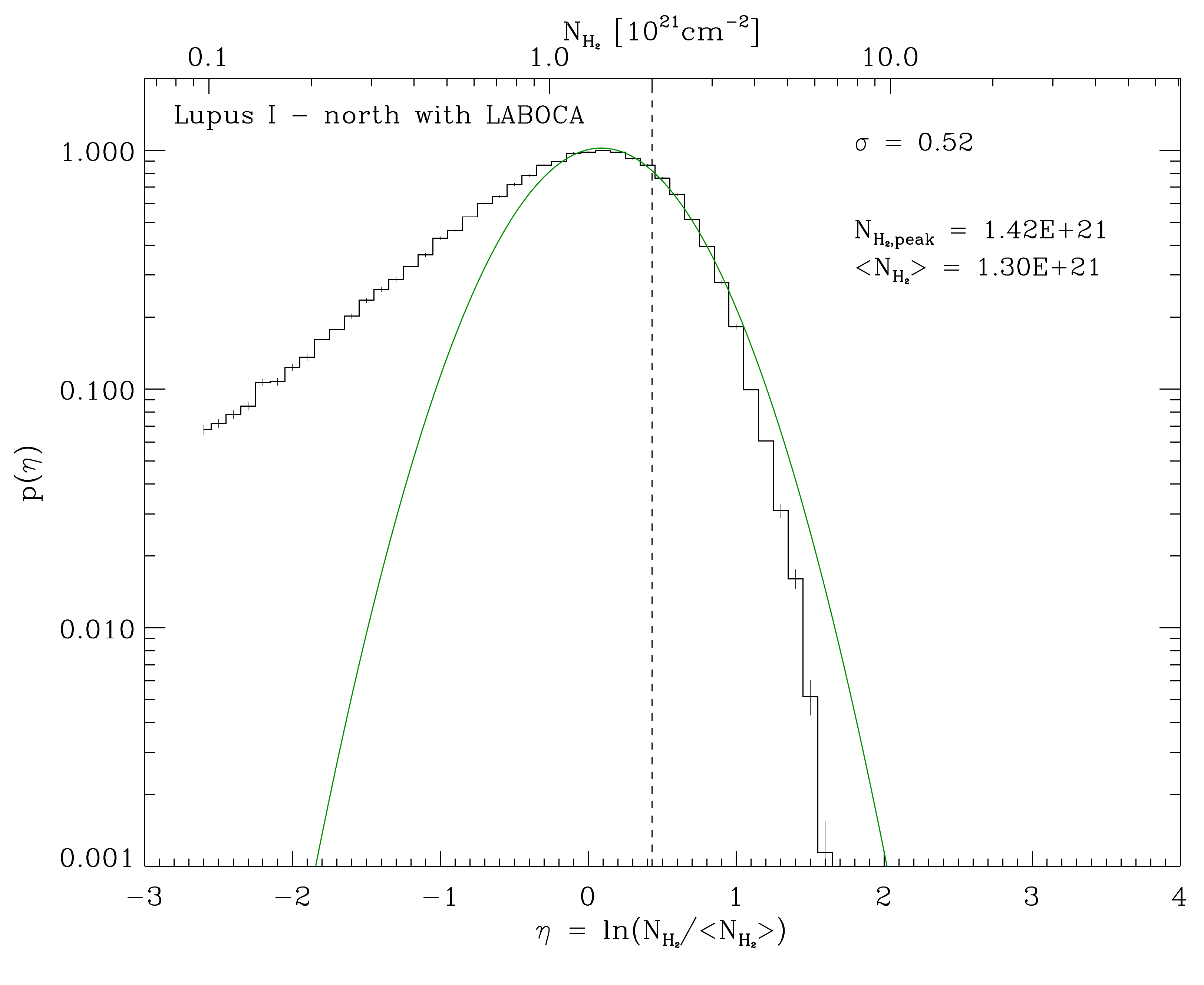}
  \label{img:Laboca_pdf_N}}
\caption[Column density PDFs of Lupus~{\rm I}.]{Column density PDFs of Lupus~{\rm I} and their model fits for the north and the center-south part of
the cloud, as well as for the entire cloud.
The error bars show the $\sqrt{N}$ uncertainties. In case of {\it Herschel} the two yellow and purple dashed lines show the fits of the two lognormals to the distribution
around the first and second peak, respectively. Their superposition is represented by the solid green line.
For LABOCA the solid green line shows the lognormal fit to the distribution around the peak. For all cases the straight solid red
line shows the power-law fit to the high-density tail of the PDF, if applicable. In each plot we give the value of the dispersion(s) of the fitted lognormal(s) $\sigma_i$,
the position(s) of the peak(s) of the lognormal(s) $\rm N_{{H_2},peak_i}$, the slope of the fitted power-law tail s, the corresponding slope of an equivalent spherical density distribution $\alpha$, the mean column density <$\rm N_\mathrm{H_2}$>, 
and the deviation point from the lognormal to the power-law tail $\rm N_{{H_2},dev}$. The vertical dashed line marks the incompleteness limit 
(lowest closed contour) which is $1.1\times10^{21}$ and $2\times10^{21}\,{\rm cm^{-2}}$ for the {\it Herschel} and LABOCA map, respectively.}
\label{img:PDFs}
\end{figure*}

We derived PDFs of the Lupus~{\rm I} cloud from the {\it Herschel} and LABOCA column density maps
for the two distinct parts of the cloud (north and center-south; see Fig.~\ref{img:herschel-col-dens-map-3d}), as well as for the whole cloud.
They are shown in Fig.~\ref{img:PDFs}. 
For a correct interpretation of the PDFs one has to consider
the completeness limit of the underlying column density map. Here we adopted the lowest closed contour
as such a limit which is $1.1\times10^{21}$ and $2\times10^{21}\,{\rm cm^{-2}}$ for the {\it Herschel} and LABOCA map, respectively.
It is marked by the vertical dashed line in the plots of Fig.~\ref{img:PDFs}.

Looking at the entire cloud the {\it Herschel} PDF is very complex with clear deviations from a simple lognormal
distribution which would be expected for a cloud that is dominated by isothermal, hydrodynamic turbulence
\citep[e.g.][]{gravitational-collaps-in-mc-klessen-2000}. The distribution shows two peaks in the low column
density regime and a power-law tail in the high density end. The first peak at $\rm N_{{H_2},peak1}=8.66\times10^{20}\,{\rm cm^{-2}}$
falls below our completeness limit, so it might be just reflecting the drop of observational sensitivity and will not be used
for interpretation.
But the distribution above the limit can be well represented by a fit of
two lognormals around the first and second peak which is at $\rm N_{{H_2},peak2}=2.57\times10^{21}\,{\rm cm^{-2}}$.
The width of the first component is $\sigma_1=0.56$, but could also be broader due to the possible underestimate
of the column densities below the completeness limit. Nevertheless, it is more than twice as broad as the width
of the second lognormal ($\sigma_2=0.26$) and larger than in other nearby clouds like Maddalena, 
Auriga \citep{pdf-column-density-orion-schneider-2013} or Aquila \citep{pdf-column-density-schneider-2015}.
This can be a sign of broadening by turbulence and external compressive 
forcing \citep[see][]{turbulence-ism-forcing-federrath-2010,pdf-hii-turbulence-shock-simulation-tremblin-2012,
sf-efficiency-in-turbulent-magnetized-clouds-federrath-klessen-2013}.
The power-law tail follows a slope of $\rm s=-2.55$ or corresponding $\alpha=1.78$ for a spherical density distribution.
The deviation from the lognormal into the power-law occurs near $\rm N_{{H_2},dev}=3.6\times10^{21}\,{\rm cm^{-2}}$ which 
is slightly below the recent values by \cite{pdf-column-density-schneider-2015} who found the transition into the power-law tail
to be at $A_V\sim4-5$~mag\footnote{$\rm N(H_2)/A_v=0.94\times10^{21}\,cm^{-2}\,mag^{-1}$ \citep{nh2-to-av-factor-bohlin-1978}.} 
in their investigation of four low-mass and high-mass star forming regions. But it is higher than the result of 
\cite{pdf-serpens-core-roccatagliata-2015} who find the deviation to occur at $A_V\sim1.2-2$~mag in the Serpens Core region.

From the {\it Herschel} data the center-south part of the cloud shows a PDF very similar to that of the entire cloud. Here again two peaks are present
and the distribution above the completeness limit can be fitted by a double lognormal
and a power-law tail. The width of the first fitted lognormal is narrower, the width of the second broader than
those of the entire cloud with values of $\sigma_1=0.46$ and $\sigma_2=0.35$. The slope of the power-law is slightly flatter with $\rm s=-2.43$, but leading
to a comparable exponent $\alpha=1.82$.

Also in the northern part of the cloud the double peak lognormal profile of the PDF from {\it Herschel} is found again.
But now there is no excess above the second lognormal at higher densities.
The peak position of the first lognormal is lower ($\rm N_{{H_2},peak1}=7.94\times10^{20}\,{\rm cm^{-2}}$) compared
to the center-south and the whole cloud, but with similar width ($\sigma_1=0.55$).
The peak position of the second lognormal ($\rm N_{{H_2},peak2}=2.23\times10^{21}\,{\rm cm^{-2}}$) is lower than 
that of the whole cloud and its width ($\sigma_2=0.18$) is approximately half as broad.

Although it is not possible to strictly constrain the shape of the PDF below the completeness limit,
one can say that the distribution above shows not just two components (lognormal and a power-law) but 
at least three because of the second peak that appears before
the deviation into the power-law. Such a shape could be attributed to an initial turbulent cloud
which later was compressed by an external driving agent. Such an agent can either be an ionization
front, an expanding shell driven by winds or/and supernovae or colliding flows. The initial lognormal
form of the PDF of the cloud develops a second component caused by the dense compressed gas.
This behavior was found in simulations by \cite{pdf-hii-turbulence-shock-simulation-tremblin-2012} 
of an initial turbulent medium that is ionized and heated by an ionization source. They concluded that a 
double peak is present when the ionized gas pressure dominates over the turbulent ram pressure of the cloud.
\cite{pdf-double-peaks-colliding-flows-matsumoto-2015} studied the evolution of turbulent molecular clouds swept by a
 colliding flow. They found that the PDF exhibits two peaks in the case when the Mach numbers of the initial turbulence
and the colliding flow are of the same order $M=10$ (model HT10F10; see their Fig.~11b) and the line of sight is perpendicular
to the colliding flow. Then the low column density peak represents the colliding flow and the
higher column density peak the sheet cloud, respectively.
Observationally, double peak PDFs have been reported and studied recently by \cite{pdf-rosette-cluster-formation-Schneider-2012},
\cite{pdf-auriga-california-herschel-harvey-2013}, and \cite{herschel-hii-compression-pdf-tremblin-2014} 
for several nearby clouds exposed to an ionization source.
Areas of the cloud close to the ionization front indeed showed the predicted second peak in the PDF due to
the compression induced by the expansion of the ionized gas into the molecular cloud.
In the case of Lupus~{\rm I} the source of the compression is very likely the expanding H\,{\rm I} shell around
the USco sub-group of Sco-Cen 
(possibly together with a supernova that exploded within USco) 
and the wind bubble of the remaining B-stars of UCL pushing from the eastern and western
side of the cloud, respectively. This scenario will be discussed in more detail in Sec.~\ref{sec:surrounding}.

Lognormal PDFs indicate shock waves \citep{shocks-ism-imf-kevlahan-2009} or turbulence 
\citep[e.g.][]{structure-flows-ism-vazquez-1994,turbulence-ism-forcing-federrath-2010}.
\citet{pdf-hii-turbulence-shock-simulation-tremblin-2012} have shown in a simulation of Stroemgren spheres advancing
into turbulent regions that the region compressed by the ionization front also has a lognormal density PDF,
but shifted to higher density due to the compression. A PDF of a region that includes compressed and undisturbed parts
of the cloud will thus show a double peak PDF.

The power-law tail that is seen in our PDFs could be explained by active star formation and the transition
to a gravity dominated density regime represented by the star forming cores.
This can be shown by comparing the PDFs of the two regions in the north
and the center-south. In the north where there is almost
no star formation and just one pre-stellar core and $\approx20$ unbound cores can be found \citep{lupus-clouds-herschel-rygl-2013},
the PDF shows no power-law tail in the high density regime. In the center-south instead where almost all the star
formation activity takes place with plenty of very young bound cores, the PDF shows the power-law tail very clearly. 
Also numerical studies have shown that the PDF for an actively star forming region develops a clear deviation from the lognormal
in form of a power-law tail \citep{pdf-evolution-gravity-turbulence-ballesteros-2011,pdf-column-density-evolution-numerical-girichidis-2014}.

With LABOCA the PDF of Lupus~{\rm I} can be modeled as one lognormal with a power-law tail at high densities
($\rm N_{{H_2},dev}=6.0\times10^{21}\,{\rm cm^{-2}}$). The peak of the lognormal distribution lies 
at $\rm N_{{H_2},peak}=1.43\times10^{21}\,{\rm cm^{-2}}$ which is below the completeness limit of the
LABOCA column density map. The width of the lognormal is broad ($\sigma=0.56$), but uncertain due to the peak position
and thus possibly even broader. The power-law slope ($s=-2.66$, $\alpha=1.75$) is very similar to the one of the {\it Herschel} PDF indicating
the star formation activity in the cloud.

The PDF of the center-south region shows the same behavior as the one for the whole cloud. The lognormal
part is slightly broader ($\sigma=0.60$), but the power-law slope ($s=-2.59$, $\alpha=1.77$) and the 
deviation point ($6.0\times10^{21}\,{\rm cm^{-2}}$) are almost the same compared to the whole cloud.
In the northern part only the now narrower lognormal is seen instead ($\sigma=0.52$),
indicating the lack of star formation activity
and the dominance by the turbulence induced trough the compression.
In all three cases the positions of the peak are very close to each other.

As was already seen in the column density map from LABOCA, its PDF shows only
the dense cores and the intercore medium. The instrument is not sensitive to
all material that can still be probed with {\it Herschel}.
With LABOCA we thus sample more the second lognormal of the {\it Herschel} PDF and the power-law tail.

\subsection{Core distribution}
\label{sec:core-distribution}

For the core analysis in the LABOCA map we used the {\tt Clumpfind} package \citep{clumpfind}.
It decomposes the emission of the map into a set of clumps or cores contouring the data
at given threshold levels.
The results can be found in Tab.~\ref{tbl:laboca-cores} and the distribution of the cores (represented by the white crosses)
is shown in Fig.~\ref{img:laboca-map}. The algorithm identified 15 different cores
with masses between 0.07 and $1.71\,M_\odot$. Their total mass is $8.37\,M_\odot$ and their total flux
is 25.39~Jy. This corresponds to $\approx5\%$ of the total mass of Lupus~{\rm I} (see Sec.~\ref{sec:total-mass-lupus}) and $\approx5\%$ 
of the total flux of the LABOCA map, respectively. For the computation of the core masses, we derived their
temperatures from the {\it Herschel} SPIRE SED fit temperature map as a mean temperature
within the ellipse representing the core. With this temperature the mass of the cores was calculated following \cite{atlasgal}

\begin{equation}
M=\frac{d^2 F_\nu R}{B_\nu(T_\mathrm{d}) \kappa_\nu}
\end{equation}
where $d$ is the distance to Lupus~{\rm I}, $F_\nu$ the total flux of the core, $R=100$ the dust-to-gas ratio,
and $\kappa_\nu = 1.32\,{\rm cm}^2\,{\rm g}^{-1}$ (see Sec.~\ref{sec:col-dens-map-Laboca}).
Most interestingly, all but one core are found in the center-south part of Lupus~{\rm I}. This confirms our interpretation of the
column density maps that is the part where active star formation is going.

The distribution of the cores from the {\it Herschel} data \citep[][]{lupus-clouds-herschel-rygl-2013}
agrees with this picture and the core distribution from our LABOCA map. The northern part of Lupus~{\rm I}
is mainly populated by unbound cores and just one pre-stellar core wheras the center-south is dominated
by pre-stellar cores. Many of those coincide with our LABOCA cores.
However, a direct assignment is not easily possible since the {\it Herschel} coordinates
can only be estimated from the source images in the paper of \cite{lupus-clouds-herschel-rygl-2013} 
and in some cases several {\it Herschel} sources seem to be on the position
of one LABOCA source or vice-versa.

\begin{figure}[htb]
 \centering
 \includegraphics[width=0.35\textwidth, keepaspectratio]{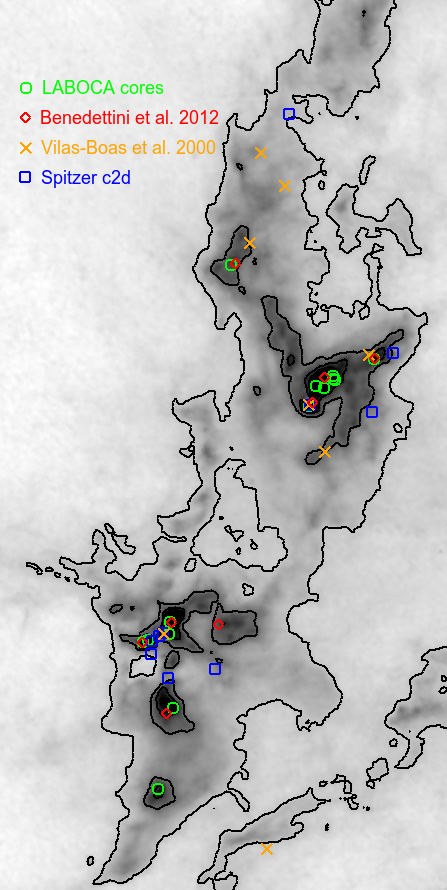}
 \caption{{\it Herschel} column density map of the center-south region with contour levels of 1.1, 4, 7, and $10\times10^{21}\,{\rm
 cm^{-2}}$. The different symbols mark the center positions of the cores found with LABOCA and of objects found in other studies as
 discussed in Section~\ref{sec:core-distribution}.}
\label{img:c2d-ysos-on-herschel}
\end{figure}

\cite{lupus-clouds-spec-cores-Benedettini-2012} found eight dense cores in Lupus~{\rm I} using high-density
molecular tracers at 3 and 12~mm with the Mopra telescope (red diamonds in Figure~\ref{img:c2d-ysos-on-herschel}).
Seven of those have one or more 
counterparts in our LABOCA map. The matches are given in Table~\ref{tbl:laboca-cores}. 
They classify five of their cores as very young protostars or pre-stellar
cores (Lup1 C1-C3, C5, and C8) and the remaining three also as very likely to be protostellar or pre-stellar. 
For three of their cores (Lup1 C4, C6, and C7) they calculated a kinetic temperature of $\approx12$~K.
This agrees within 20\% with the core temperatures derived from the {\it Herschel} SPIRE SED fit temperature map.

From the 17 young stellar objects found in Lupus~{\rm I} by the {\it Spitzer} c2d near-infrared 
survey \citep{lupus-clouds-spitzer-c2d-chapman-2007,lupus-clouds-spitzer-c2d-merin-2008}
11 lie within the boundaries of the LABOCA map (blue boxes in Figure~\ref{img:c2d-ysos-on-herschel}).
But only one object (IRAS~15398--3359) clearly matches one
of our LABOCA cores (\#1). This is the long known low-mass class~0 protostar 
IRAS~15398--3359 (inside the B228 core) which has a molecular outflow 
\citep[see recently e.g.][]{lupusI-iras-outflow-oya-2014,outflows-in-low-mass-ps-Dunham-2014}.
Two other objects of the c2d survey are close, but offset by $\approx0.5\arcmin$ to the center of our
cores \#3 and \#5, respectively. Nevertheless, each of those two objects still lies within the boundaries of 
the corresponding ellipse representing the LABOCA core (see Table~\ref{tbl:laboca-cores}).
Most of the LABOCA cores are potentially at a very early evolutionary stage, 
i.e. without a protostar inside to heat it and eventually destroy the surrounding dust
envelope to allow near-infrared radiation to escape. In fact the only cores with a $70\,\mu$m counterpart,
which is a good proxy for a protostar inside a core, are number \#1 and \#3.
Therefore, we do not expect to find many {\it Spitzer} counterparts. 

\cite{lupus-clouds-13co-cores-vilas-boas-2000} found 15 (14) condensations in $\rm C^{18}O$ ($\rm^{13}CO$) with
the 15~m SEST telescope. Eight cores lie
within our LABOCA map (orange crosses in Figure~\ref{img:c2d-ysos-on-herschel}) 
and of those three (Lu7, Lu10, B228) coincide with LABOCA detected cores (\#12, \#5, and \#1).
Their $\rm C^{18}O$ excitation temperatures are 9, 11, and 10~K, respectively. This is $\approx20-40\%$ lower than
the temperatures from the {\it Herschel} map.
For cores Lu7 and Lu10 they calculated a mass from $\rm C^{18}O$ of 11.7 and 3.1~$M_\odot$, respectively. This is
a factor of $\approx20$ and $\approx7$ higher than the LABOCA masses. But the sizes of their condensations were
on average larger than at least three times their beam size of $48\arcsec$.

Figure~\ref{img:c2d-ysos-on-herschel} shows the {\it Herschel} column density map of the center-south region
with contour levels of 1.1, 4, 7, and $10\times10^{21}\,{\rm cm^{-2}}$. Overploted are the LABOCA cores found
in this study (green circles), as well as the above mentioned cores and YSOs found by \cite{lupus-clouds-spec-cores-Benedettini-2012}
(red diamonds), \cite{lupus-clouds-13co-cores-vilas-boas-2000} (orange crosses), and the {\it Spitzer} c2d survey (blue boxes), respectively.

Our findings confirm that Lupus~{\rm I} harbors a population of very young cores that
are forming stars just now and probably have ages below 1~Myr. This could support the idea of
an external shock agent, like the USco shell, sweaping up the cloud and triggering the simultaneous
formation of new stars within the cloud.

\begin{table*}[htb]
 \centering
 \caption[]{Parameters of the 15 cores detected in the LABOCA map with {\tt Clumpfind} (see Fig.~\ref{img:laboca-map}).
The beam size (HPBW) and area used to convert intensities into fluxes is
 $21.2\arcsec$ and $509~{\rm arcsec^2}$, respectively. The total mass of all cores is $\approx 8\,M_\odot$. The 4th and 5th column give the FWHM of the
ellipses representing the cores. The 6th, 7th, and 8th column give the core's peak intensity, total flux, and mass, respectively. The core temperature shown 
in the 9th column was derived from the {\it Herschel} SPIRE SED fit temperature map as a mean temperature
within the ellipse representing the core. The last column gives the matches to dense cores found by \cite{lupus-clouds-spec-cores-Benedettini-2012}.}
 \begin{tabular}{r c c c c c c c c c}
 \hline\hline
 \noalign{\smallskip}
  \# &    Ra &  Dec  &    FWHM$_{\rm x}$  &  FWHM$_{\rm y}$  &  I$_{\rm max}$ &  F$_{\rm tot}$ &   M$_{\rm core}$ &   T$_{\rm core}$ &   Counterpart \\   &    &     &   $[\arcsec]$  &   $[\arcsec]$  & [Jy/beam]  &   [Jy]   &   [M$_\odot$] &   [K]  &         \\
 \noalign{\smallskip}
 \hline
 \noalign{\smallskip}
   1 &   15:43:01.68 &  $-$34:09:08.9 &  58.22  &  44.38  &    1.37 &    5.61  &    1.71 &   14.0  &   Lup1 C4 \\    
   2 &   15:44:59.85 &  $-$34:17:08.8 &  62.51  &  44.09  &    0.51 &    3.97  &    1.59 &   12.0  &   Lup1 C6 \\  
   3 &   15:45:13.58 &  $-$34:17:08.8 &  46.56  &  33.45  &    0.33 &    1.48  &    0.40 &   15.0  &   Lup1 C7 \\    
   4 &   15:45:16.02 &  $-$34:16:56.6 &  35.31  &  34.95  &    0.31 &    1.00  &    0.27 &   15.0  &   Lup1 C7 \\    
   5 &   15:45:03.80 &  $-$34:17:57.3 &  57.22  &  42.24  &    0.28 &    1.41  &    0.46 &   13.5  &   Lup1 C6 \\    
   6 &   15:42:52.86 &  $-$34:08:02.1 &  96.42  &  49.94  &    0.28 &    3.85  &    1.34 &   13.0  &   Lup1 C3 \\    
   7 &   15:42:43.57 &  $-$34:08:14.2 &  70.62  &  44.38  &    0.25 &    1.64  &    0.57 &   13.0  &   Lup1 C3 \\    
   8 &   15:42:44.55 &  $-$34:08:32.5 &  30.55  &  20.85  &    0.25 &    0.48  &    0.17 &   13.0  &   Lup1 C3 \\    
   9 &   15:42:44.06 &  $-$34:08:44.6 &  36.82  &  19.26  &    0.24 &    0.49  &    0.17 &   13.0  &   Lup1 C3 \\    
  10 &   15:42:50.42 &  $-$34:08:38.5 &  53.28  &  30.34  &    0.22 &    1.05  &    0.36 &   13.0  &   Lup1 C3 \\    
  11 &   15:45:25.10 &  $-$34:24:01.8 &  42.46  &  69.57  &    0.22 &    1.63  &    0.53 &   13.5  &   Lup1 C8 \\    
  12 &   15:42:23.02 &  $-$34:09:33.2 &  68.97  &  46.08  &    0.21 &    1.72  &    0.53 &   14.0  &   Lup1 C1 \\    
  13 &   15:39:09.92 &  $-$33:25:30.9 &  27.24  &  27.70  &    0.20 &    0.34  &    0.07 &   17.0  &   -- \\    
  14 &   15:45:55.75 &  $-$34:29:23.7 &  28.10  &  26.28  &    0.20 &    0.29  &    0.08 &   14.5  &   -- \\    
  15 &   15:42:47.90 &  $-$33:53:15.3 &  43.80  &  19.59  &    0.19 &    0.43  &    0.12 &   15.0  &   Lup1 C2 \\   
 \noalign{\smallskip}
 \hline
 \end{tabular}
\label{tbl:laboca-cores}
\end{table*}

\subsection{Total mass estimates of Lupus~{\rm I}}
\label{sec:total-mass-lupus}

To derive the total mass of the cloud from the three different
column density maps we defined a polygon around
the filament (delineated in Fig.~\ref{img:laboca-map} on the LABOCA map)
to derive the total mass always in the same area ($\rm \approx1\,deg^2$).
The total gas and dust mass was then calculated via the formula

\begin{equation}
\rm M^{\rm tot}_{Lupus~{\rm I}}=\sum\,N_\mathrm{H_2}\,\mu_\mathrm{H_2}\,m_H\,A_p
\end{equation}
with $A_p$ the area of a pixel in $\rm cm^2$. As the common lower level we chose $\rm N_\mathrm{H_2}>10^{21}\,cm^{-2}$ which corresponds to $\rm A_V>1$~mag.
The resulting total mass for Lupus~{\rm I} is $\rm M^{\rm tot}_{Lupus~{\rm I}}\approx171\,M_\odot$, $\approx174\,M_\odot$, and $\approx164\,M_\odot$
for the {\it Planck}, {\it Herschel}, and LABOCA data. 
This means that the total masses calculated from the three data sets agree with each other.
Comparing the total mass of the cloud to the total mass in cores from LABOCA ($\approx8\,M_\odot$) one sees that only about 5\% of the mass
is concentrated in the densest condensations. Most of the dust and gas is in more diffuse components.

The most recent literature value for the total mass of Lupus~{\rm I} is 
from the {\it Herschel} data by \cite{lupus-clouds-herschel-rygl-2013}.
They calculated the total mass in a much bigger area than this work ($\approx 4.5\,{\rm deg}^2$ compared to our $\approx 1\,{\rm deg}^2$)
and used a different method to create their column density map as mentioned already
in Sec.~\ref{sec:results:Column density and temperature maps}.
Therefore, a difference of about a factor $4-5$ arises between their value of
$M=830\,M_\odot$ (for $A_V>2$~mag) and our finding.

Other literature values cover a wide range of total masses
for Lupus~{\rm I}, depending on the tracer used and the size of the area
that was considered.
From the {\it Spitzer} c2d near-infrared extinction
maps \cite{lupus-clouds-spitzer-c2d-merin-2008} determined
a total mass of $479\,M_\odot$ 
for $A_V>3$. Various CO measurements \citep[e.g.][]{lupus-clouds-13co(1-0)-Tachihara-1996,
lupus-clouds-c18o-Hara-1999,lupus-clouds-13co(2-1)-Tothill-2009} yielded
values of $\approx280 - 880\,M_\odot$. Direct comparisons with our values are not
always possible since all maps cover different parts of the Lupus~{\rm I} cloud complex and the
material is not homogeneously distributed to allow scaling the mass with the area.
But we note that our values agree with most of the literature
values within a factor of 2-3 which is expected considering the uncertainties
in the choice of the dust model, the dust-to-gas ratio and the CO-to-H$_2$ conversion factors.

\section{The surroundings of Lupus~{\rm I} and the interaction with USco and UCL}
\label{sec:surrounding}

\begin{figure*}[htb]
\centering
 \includegraphics[width=\linewidth, keepaspectratio]{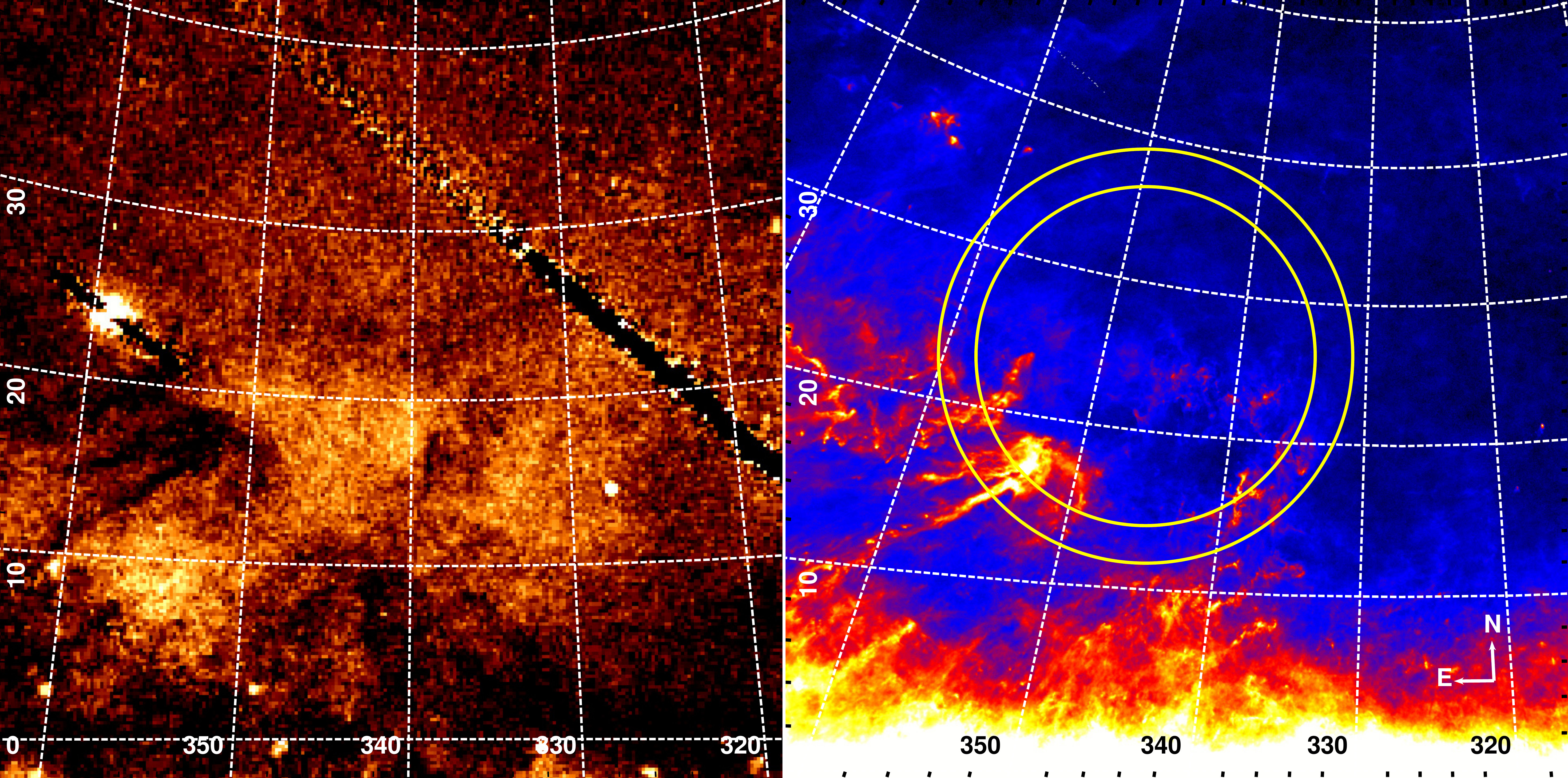} 
\caption{The surrounding of Lupus~{\rm I} in X-rays (ROSAT $3/4$~keV, left) and dust emission ({\it Planck} $850\,\mu$m, right). There are two
X-ray bubbles on either side of the Lupus~{\rm I} ridge and both Lupus~{\rm I} and $\rho$~Ophiuchus as well as the little cloudlets north of them seen in dust emission
have X-ray shadows. The yellow dashed circles in the right image give the position of the inner and outer edge of the USco H\,{\rm I} shell determined from our preliminary
analysis of the GASS H\,{\rm I} data. The center lies at $l=347\degr$, $b=25\degr$ and the inner and outer radius is $12\degr$ and $15\degr$, respectively.}
\label{img:lupus-usco-xray-planck}
\end{figure*}

\begin{figure*}[htb]
\centering
 \includegraphics[width=\linewidth, keepaspectratio]{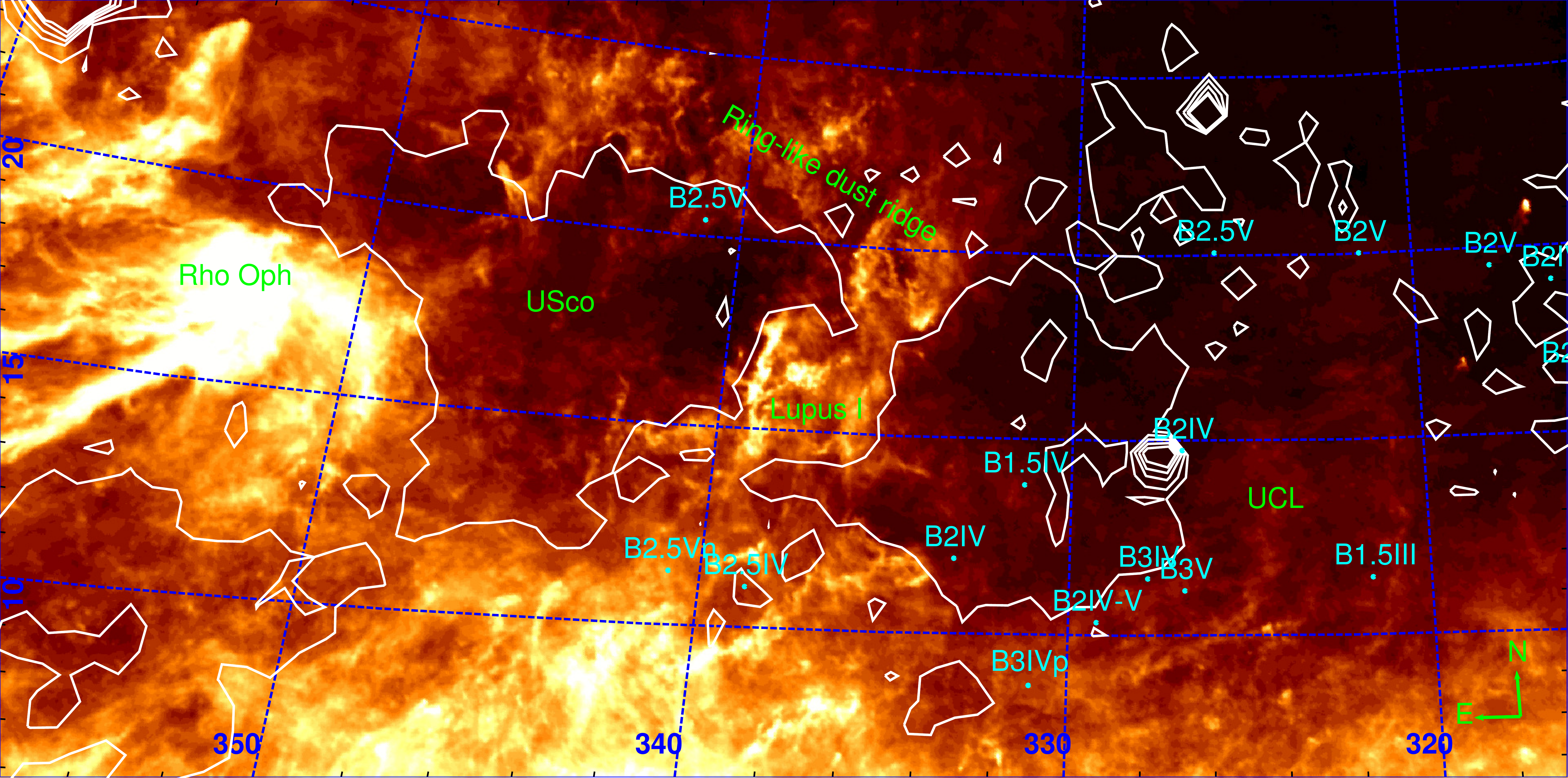} 
\caption{The surroundings of Lupus~{\rm I} seen in dust emission ({\it Planck} $350\,\mu$m). The white contours mark the X-ray emission (ROSAT 0.8~keV)
from $0.27-1\times10^{-3}$\,counts/s. The cyan dots mark the positions of the B1-B3 stars of UCL from the {\it Hipparcos} catalog. Labels are given in green.}
\label{img:lupus-usco-xray-planck-contours}
\end{figure*}

Best suited for looking at the dust surroundings of Lupus~{\rm I} are the {\it Planck} data (see Fig.~\ref{img:lupus-usco-xray-planck} and Fig.~\ref{img:lupus-usco-xray-planck-contours}). These observations at 350, 550, and $850\,\mu$m cover the whole sky at a resolution of $5\arcmin$.
Lupus~{\rm I} lies on the eastern edge (Galactic coordinates) of a ring-like dust ridge (labeled in Fig.~\ref{img:lupus-usco-xray-planck-contours})
that extends from about $b=+10\degr$ to $b=+25\degr$ in Galactic latitude with a center at about $l=+345\degr$, $b=17.5\degr$. The ridge is $\sim5\degr$ wide which
corresponds to $\sim13$~pc at the distance of Lupus~{\rm I}. Besides Lupus~{\rm I} it consists of several small molecular clouds
extending north of Lupus~{\rm I} and then bending towards the east connecting with the $\rho$~Ophiuchus molecular cloud
on the opposite site of Lupus~{\rm I}. Further west of Lupus~{\rm I} no dust emission
is seen in the {\it Planck} maps. The same is true for the inside of the dust ridge. Between $\rho$~Ophiuchus and Lupus~{\rm I}
a roundish dust void is seen.
However, these two dust voids on either side of Lupus~{\rm I} are filled with hot X-ray gas which can be seen
with ROSAT (left in Fig.~\ref{img:lupus-usco-xray-planck}). 
Fig.~\ref{img:lupus-usco-xray-planck-contours}
shows the dust emission seen by {\it Planck} in $350\,\mu$m. Overlayed in white are the contours of the ROSAT
diffuse X-ray emission at 3/4~keV\footnote{The very strong X-ray emission seen near the position of the B2IV star south-west of Lupus~{\rm I}
($\kappa$\,Cen; $l=326.872\degr, b=14.754\degr$) is neither related to the star nor to Sco-Cen,
but most likely caused by the Quasar [VV2006] J150255.2$-$415430.}.
Inside USco the contours follow the edge of the dust ridge indicating
that the hot X-ray gas might be in contact with the cold dust. On the western side of Lupus~{\rm I} the contours mark the outline of
the second roundish X-ray emission seen in ROSAT. The cyan dots mark the remaining early B-type stars of UCL. Both contours
seem to wrap around Lupus~{\rm I} what might be a sign that the cloud is embedded in hot ISM.

These dust voids and the observed X-ray gas might be explained by the cumulative feedback of
the massive stars in the USco and UCL sub-groups of Sco-Cen. Their creation has been interpreted in a scenario of propagating 
molecular cloud formation and triggered cloud collapse and star formation taking place within the last 17~Myr in the UCL and USco sub-groups of Scorpius-Centaurus
\citep[see][]{sco-cen-HI-degeus-1992,ob-stars-triggered-sf-preibisch-zinnecker-2007}.
The expansion of the UCL H\,{\rm I} shell that started $\sim10$~Myr ago and was driven by winds of the massive stars and supernovae explosions
has probably cleared out almost all the dust and molecular material west of Lupus~{\rm I}. 
But the observed X-ray gas on that side is probably not related to the supernova explosions, because 
the X-ray luminosity of supernova-heated superbubbles dims on a timescale of typically
less than a million years via expansion losses and mixing with entrained gas \citep{feedback-superbubbles-xrays-krause-2014}.
The shell today has a radius of $\sim110$~pc \citep[as seen from H\,{\rm I} data; ][]{sco-cen-HI-degeus-1992}, generally consistent 
\citep[compare, e.g.][]{superbubble-evolution-disk-galaxies-baumgartner-breitschwerdt-2013,superbubbles-vishniac-krause-2013}
with the inferred age of the stellar group, about 17 million years. Therefore, this X-ray gas is probably 
currently heated by the remaining B-stars of UCL.
Six of those stars south-west of Lupus~{\rm I} ($325\degr<l<335\degr$ and $10\degr<b<15\degr$; 
see Fig.~\ref{img:lupus-usco-xray-planck-contours}) lie at positions that favor them as likely
being the sources of this wind bubble.

The USco H\,{\rm I} shell has started its expansion $\sim5$~Myr ago powered by the winds of the OB-stars and quite possibly a
recent supernova explosion, about 1.5~Myr ago, as suggested by the detection of the 1.8~MeV gamma ray line towards 
the USco sub-group and the detection of the pulsar PSR~J1932+1059 \citep[][and references therein]{sco-cen-al26-diehl-2010}.
What now forms the before mentioned dust ridge is probably the remaining material
of the USco parental molecular cloud that was swept up by the advancing USco shell leaving the dust void filled with X-ray gas.
From simulations, we expect large scale oscillations of the hot, X-ray emitting gas \citep{feedback-superbubbles-xrays-krause-2014},
 because the energy source is never exactly symmetric.

It seems like the UCL wind bubble could be colliding with the
USco H\,{\rm I} shell right at the position of Lupus~{\rm I} squeezing it in-between. This wind bubble might have provided a
counter-pressure to the expanding USco shell and thus favored this position for an additional compression of the shell material.
In this way a new molecular cloud could have been created there and it might explain why we do not see more very young star forming
clouds (except $\rho$~Oph) distributed within the wall of the USco shell. 
With its distance of 150~pc the cloud also is neither in the foreground of the B-stars of UCL that
have an average distance of $\sim140$~pc \citep{sco-cen-hipparcos-dezeeuw-1999} nor in front of USco \citep[$\sim145$~pc;][]{usco-full-population-preibisch-2002}.
To rule out shadow effects completely, however, a more comprehensive analysis of the X-ray data would be required.
 
From our preliminary analysis of the 
GASS H\,{\rm I} data \citep[][]{gass-hi-mcclure-griffiths-2009,gass-hi-kalberla-2010}
we see that Lupus~{\rm I} is indeed embedded within the expanding shell around USco (Kroell et al., in prep.). The shell has an expansion
velocity of $\approx7\,{\rm km\,s^{-1}}$ and a thickness of $\approx6$~pc. We estimate the current outer radius 
to be $\approx36$~pc (both inner and outer radius of the USco shell from our H\,{\rm I} model are delineated in Fig.~\ref{img:lupus-usco-xray-planck}).

These findings complement previous suggestions of an interaction of Lupus~{\rm I} (and $\rho$~Oph) with the two sub-groups of Sco-Cen 
that were originally proposed by \cite{sco-cen-HI-degeus-1992}. From the spatial and velocity structure of $\rm ^{12}CO$ observations
\cite{lupus-clouds-12co-Tachihara-2001} also found evidence for an interaction. \cite{lupus-clouds-13co(2-1)-Tothill-2009}
in their analysis of $\rm ^{13}CO$ and CO(4-3) found enhanced line widths at the western end of Lupus~{\rm I} and a velocity 
gradient across the filament, i.e. in the direction of the USco shell expansion, of $\rm \sim1\,km\,s^{-1}$ which they 
conclude to be consistent with a dynamical interaction between Lupus~{\rm I} and the USco H\,{\rm I} shell.
\cite{lupus-submm-polarimetry-matthews-2014} have found that the large-scale magnetic field is 
perpendicular to the Lupus~{\rm I} filament, i.e. points in the direction of the USco shell expansion. This might
have favored the accumulation of cold, dense atomic gas along the field lines and promote faster molecule 
formation \citep{rapid-formation-mcs-hartmann-2001,mc-formation-mhd-vazquez-semadeni-2011}.

All these are hints towards the scenario that Lupus~{\rm I} was affected by large-scale external compressing forces coming from the expansion
of the USco H\,{\rm I} shell and the UCL wind bubble.
This might explain its position, orientation and elongated shape, the appearance of a double peak in the PDF and the large population of very young pre-stellar cores
that are seen with both LABOCA and {\it Herschel}.

\section{Summary and Conclusions}
\label{sec:summary}

From our LABOCA observations of Lupus~{\rm I} as well as from archival {\it Herschel} and {\it Planck} data
we created column density maps of the cloud. In addition we calculated a temperature map from an SED fit using
the three SPIRE bands of {\it Herschel}. All maps suggest that the cloud can be divided into two distinct
regions. The northern part that has on average lower densities and higher temperatures as well as no active star
formation and the center-south part with dozens of pre-stellar cores where density and temperature
reach their maximum and minimum, respectively.

From the LABOCA and {\it Herschel} maps we derived column density PDFs for the entire cloud as well as for
the two above mentioned regions separately. The {\it Herschel} PDF of Lupus~{\rm I} showed a double peak profile
with a power-law tail. The power-law can be attributed to the star formation activity in the center-south
part of the cloud since it disappears in the PDF of the northern part. However, the double peak profile is
conserved throughout the cloud and possibly arises due to the large-scale compressions from the cumulative
massive star feedback of the Sco-Cen sub-groups.
Such a sign for compression in the PDF was previously found in both observations and simulations of
advancing ionization fronts, also supported by simulations of colliding turbulent flows. With LABOCA we probe 
the denser parts of Lupus~{\rm I} and find
one lognormal and the power-law tail for the whole cloud and the center-south. The PDF of the northern part
shows only the lognormal behavior.

The distribution of the 15 cores that were found in our LABOCA map confirms that only the center-south
part of Lupus~{\rm I} is actively forming stars whereas the north is quiescent. The cores have masses between
0.07 and $1.71\,M_\odot$ and a total mass of $\approx8\,M_\odot$. As the cloud has a total gas and dust mass
of $\approx164\,M_\odot$ (from LABOCA for $\rm N_\mathrm{H_2}>10^{21}\,cm^{-2}$) this means that $\sim5\%$ of the mass
is in cores. All of those cores are pre-stellar or protostellar confirming that we are witnessing 
a large star formation event \citep[see e.g.][]{lupus-clouds-spec-cores-Benedettini-2012,lupus-clouds-herschel-rygl-2013}.
The total mass from {\it Herschel} and {\it Planck} for $\rm N_\mathrm{H_2}>10^{21}\,cm^{-2}$ was $\approx174\,M_\odot$ and $\approx171\,M_\odot$, respectively.

We argue that the main driving agents in the formation process of Lupus~{\rm I} are the advancing USco H\,{\rm I} shell,
which most likely contains the cloud, and the UCL wind bubble 
interacting with the USco shell and hence squeezing Lupus~{\rm I} in-between these two shells in a complex manner.
The age of the population of young stellar objects suggests a compression event $\sim1-2$~Myr ago,
which could be the interaction between the two shells and possibly a supernova explosion as suggested by the gamma ray data \citep{sco-cen-al26-diehl-2010}.
This large-scale compressions might be the reason for the position, orientation and elongated shape of Lupus~{\rm I}, as well as
the double peak PDFs and the population of very young pre-stellar cores we found.

In future work we will analyze our newly performed high-resolution APEX CO line observations to study the kinematics
of the Lupus~{\rm I} cloud. We will also compare our observational findings with dedicated numerical simulations
and search for signs in the theoretical PDF that will allow us to distinguish between different scenarios
of molecular cloud formation.

\begin{acknowledgements}
We would like to thank our anonymous referee for his constructive comments which helped to improve this paper.
This work was supported by funding from Deutsche Forschungsgemeinschaft under DFG project number PR~569/10-1
in the context of the Priority Program 1573 "Physics of the Interstellar Medium".

Additional support came from funds from the Munich Cluster of Excellence "Origin and Structure of the Universe".

We thank Jean-Philippe Bernard for computing the {\it Planck} offsets for the {\it Herschel} maps.

The \textit{Herschel} spacecraft was designed, built, tested, and launched under
a contract to ESA managed by the Herschel/Planck Project team by an industrial
consortium under the overall responsibility of the prime contractor Thales
Alenia Space (Cannes), and including Astrium (Friedrichshafen) responsible for
the payload module and for system testing at spacecraft level, Thales Alenia
Space (Turin) responsible for the service module, and Astrium (Toulouse) responsible
for the telescope, with in excess of a hundred subcontractors.

Based on observations obtained with {\it Planck} (\url{http://www.esa.int/Planck}), an ESA science mission with instruments and contributions 
directly funded by ESA Member States, NASA, and Canada.

HIPE is a joint development by the {\it Herschel} Science Ground Segment Consortium, consisting of ESA, the NASA {\it Herschel} Science Center, and the HIFI, PACS and SPIRE consortia.

This research has made use of the SIMBAD database,
operated at CDS, Strasbourg, France.

We acknowledge the use of NASA's {\it SkyView} facility (http://skyview.gsfc.nasa.gov) located at NASA Goddard Space Flight Center.

\end{acknowledgements}

\bibliographystyle{aa} 
\bibliography{lupuspaper_v11_final_16_09_2015_literatur} 

\begin{thebibliography}{80}
\expandafter\ifx\csname natexlab\endcsname\relax\def\natexlab#1{#1}\fi

\bibitem[{{Alves de Oliveira} {et~al.}(2014){Alves de Oliveira}, {Schneider},
  {Mer{\'{\i}}n}, {Prusti}, {Ribas}, {Cox}, {Vavrek}, {K{\"o}nyves},
  {Arzoumanian}, {Puga}, {Pilbratt}, {K{\'o}sp{\'a}l}, {Andr{\'e}}, {Didelon},
  {Men'shchikov}, {Royer}, {Waelkens}, {Bontemps}, {Winston}, \&
  {Spezzi}}]{herschel-chamaeleon-sed-fit-oliveira-2014}
{Alves de Oliveira}, C., {Schneider}, N., {Mer{\'{\i}}n}, B., {et~al.} 2014,
  A\&A, 568, A98

\bibitem[{{Andr{\'e}} {et~al.}(2014){Andr{\'e}}, {Di Francesco},
  {Ward-Thompson}, {Inutsuka}, {Pudritz}, \&
  {Pineda}}]{filaments-sf-ppvi-review-andre-2014}
{Andr{\'e}}, P., {Di Francesco}, J., {Ward-Thompson}, D., {et~al.} 2014,
  Protostars and Planets VI, 27

\bibitem[{{Andr{\'e}} {et~al.}(2010){Andr{\'e}}, {Men'shchikov}, {Bontemps},
  {K{\"o}nyves}, {Motte}, {Schneider}, {Didelon}, {Minier}, {Saraceno},
  {Ward-Thompson}, {di Francesco}, {White}, {Molinari}, {Testi}, {Abergel},
  {Griffin}, {Henning}, {Royer}, {Mer{\'{\i}}n}, {Vavrek}, {Attard},
  {Arzoumanian}, {Wilson}, {Ade}, {Aussel}, {Baluteau}, {Benedettini},
  {Bernard}, {Blommaert}, {Cambr{\'e}sy}, {Cox}, {di Giorgio}, {Hargrave},
  {Hennemann}, {Huang}, {Kirk}, {Krause}, {Launhardt}, {Leeks}, {Le Pennec},
  {Li}, {Martin}, {Maury}, {Olofsson}, {Omont}, {Peretto}, {Pezzuto}, {Prusti},
  {Roussel}, {Russeil}, {Sauvage}, {Sibthorpe}, {Sicilia-Aguilar}, {Spinoglio},
  {Waelkens}, {Woodcraft}, \& {Zavagno}}]{herschel-gould-belt-andree-2010}
{Andr{\'e}}, P., {Men'shchikov}, A., {Bontemps}, S., {et~al.} 2010, A\&A, 518,
  L102

\bibitem[{{Aniano} {et~al.}(2011){Aniano}, {Draine}, {Gordon}, \&
  {Sandstrom}}]{convolution-kernels-aniano-2011}
{Aniano}, G., {Draine}, B.~T., {Gordon}, K.~D., \& {Sandstrom}, K. 2011, PASP,
  123, 1218

\bibitem[{{Ballesteros-Paredes} {et~al.}(2011){Ballesteros-Paredes},
  {V{\'a}zquez-Semadeni}, {Gazol}, {Hartmann}, {Heitsch}, \&
  {Col{\'{\i}}n}}]{pdf-evolution-gravity-turbulence-ballesteros-2011}
{Ballesteros-Paredes}, J., {V{\'a}zquez-Semadeni}, E., {Gazol}, A., {et~al.}
  2011, MNRAS, 416, 1436

\bibitem[{{Banerjee} {et~al.}(2009){Banerjee}, {V{\'a}zquez-Semadeni},
  {Hennebelle}, \& {Klessen}}]{clumps-mhd-mc-formation-banerjee-2009}
{Banerjee}, R., {V{\'a}zquez-Semadeni}, E., {Hennebelle}, P., \& {Klessen},
  R.~S. 2009, MNRAS, 398, 1082

\bibitem[{{Battersby} {et~al.}(2014){Battersby}, {Bally}, {Dunham}, {Ginsburg},
  {Longmore}, \&
  {Darling}}]{herschel-physical-dust-properties-massive-sf-regions-battersby-2014}
{Battersby}, C., {Bally}, J., {Dunham}, M., {et~al.} 2014, ApJ, 786, 116

\bibitem[{{Baumgartner} \&
  {Breitschwerdt}(2013)}]{superbubble-evolution-disk-galaxies-baumgartner-breitschwerdt-2013}
{Baumgartner}, V. \& {Breitschwerdt}, D. 2013, A\&A, 557, A140

\bibitem[{{Belloche} {et~al.}(2011){Belloche}, {Schuller}, {Parise},
  {Andr{\'e}}, {Hatchell}, {J{\o}rgensen}, {Bontemps}, {Wei{\ss}}, {Menten}, \&
  {Muders}}]{chamaeleon-laboca-belloche-2011}
{Belloche}, A., {Schuller}, F., {Parise}, B., {et~al.} 2011, A\&A, 527, A145

\bibitem[{{Benedettini} {et~al.}(2012){Benedettini}, {Pezzuto}, {Burton},
  {Viti}, {Molinari}, {Caselli}, \&
  {Testi}}]{lupus-clouds-spec-cores-Benedettini-2012}
{Benedettini}, M., {Pezzuto}, S., {Burton}, M.~G., {et~al.} 2012, MNRAS, 419,
  238

\bibitem[{{Bernard} {et~al.}(2010){Bernard}, {Paradis}, {Marshall}, {Montier},
  {Lagache}, {Paladini}, {Veneziani}, {Brunt}, {Mottram}, {Martin},
  {Ristorcelli}, {Noriega-Crespo}, {Compi{\`e}gne}, {Flagey}, {Anderson},
  {Popescu}, {Tuffs}, {Reach}, {White}, {Benedettini}, {Calzoletti},
  {Digiorgio}, {Faustini}, {Juvela}, {Joblin}, {Joncas}, {Mivilles-Deschenes},
  {Olmi}, {Traficante}, {Piacentini}, {Zavagno}, \&
  {Molinari}}]{dust-temperature-herschel-planck-bernard-2010}
{Bernard}, J.-P., {Paradis}, D., {Marshall}, D.~J., {et~al.} 2010, A\&A, 518,
  L88

\bibitem[{{Blaauw}(1964)}]{sco-cen-blaauw-1964}
{Blaauw}, A. 1964, in IAU Symposium, Vol.~20, The Galaxy and the Magellanic
  Clouds, ed. F.~J. {Kerr}, 50

\bibitem[{{Bohlin} {et~al.}(1978){Bohlin}, {Savage}, \&
  {Drake}}]{nh2-to-av-factor-bohlin-1978}
{Bohlin}, R.~C., {Savage}, B.~D., \& {Drake}, J.~F. 1978, ApJ, 224, 132

\bibitem[{{Cambr{\'e}sy}(1999)}]{lupus-extinction-star-counts-cambresy-1999}
{Cambr{\'e}sy}, L. 1999, A\&A, 345, 965

\bibitem[{{Chapman} {et~al.}(2007){Chapman}, {Lai}, {Mundy}, {Evans}, {Brooke},
  {Cieza}, {Spiesman}, {Rebull}, {Stapelfeldt}, {Noriega-Crespo}, {Lanz},
  {Allen}, {Blake}, {Bourke}, {Harvey}, {Huard}, {J{\o}rgensen}, {Koerner},
  {Myers}, {Padgett}, {Sargent}, {Teuben}, {van Dishoeck}, {Wahhaj}, \&
  {Young}}]{lupus-clouds-spitzer-c2d-chapman-2007}
{Chapman}, N.~L., {Lai}, S.-P., {Mundy}, L.~G., {et~al.} 2007, ApJ, 667, 288

\bibitem[{{Comer{\'o}n}(2008)}]{lupus-clouds-sfhb}
{Comer{\'o}n}, F. 2008, Handbook of Star Forming Regions, Volume II, ed.
  {Reipurth, B.}, 295

\bibitem[{{Dawson}(2013)}]{supershells-mc-formation-review-dawson-2013}
{Dawson}, J.~R. 2013, PASA, 30, 25

\bibitem[{{Dawson} {et~al.}(2011){Dawson}, {McClure-Griffiths}, {Dickey}, \&
  {Fukui}}]{mcs-in-supershells-dawson-2011}
{Dawson}, J.~R., {McClure-Griffiths}, N.~M., {Dickey}, J.~M., \& {Fukui}, Y.
  2011, ApJ, 741, 85

\bibitem[{{Dawson} {et~al.}(2015){Dawson}, {Ntormousi}, {Fukui}, {Hayakawa}, \&
  {Fierlinger}}]{gmc-formation-colliding-shells-dawson-2015}
{Dawson}, J.~R., {Ntormousi}, E., {Fukui}, Y., {Hayakawa}, T., \& {Fierlinger},
  K. 2015, ApJ, 799, 64

\bibitem[{{de Bruijne}(1999)}]{sco-cen-stars-hipparcos-debruijne-1999}
{de Bruijne}, J.~H.~J. 1999, MNRAS, 310, 585

\bibitem[{{de Geus}(1992)}]{sco-cen-HI-degeus-1992}
{de Geus}, E.~J. 1992, A\&A, 262, 258

\bibitem[{{de Zeeuw} {et~al.}(1999){de Zeeuw}, {Hoogerwerf}, {de Bruijne},
  {Brown}, \& {Blaauw}}]{sco-cen-hipparcos-dezeeuw-1999}
{de Zeeuw}, P.~T., {Hoogerwerf}, R., {de Bruijne}, J.~H.~J., {Brown}, A.~G.~A.,
  \& {Blaauw}, A. 1999, AJ, 117, 354

\bibitem[{{Diehl} {et~al.}(2010){Diehl}, {Lang}, {Martin}, {Ohlendorf},
  {Preibisch}, {Voss}, {Jean}, {Roques}, {von Ballmoos}, \&
  {Wang}}]{sco-cen-al26-diehl-2010}
{Diehl}, R., {Lang}, M.~G., {Martin}, P., {et~al.} 2010, A\&A, 522, A51

\bibitem[{{Dobbs} {et~al.}(2014){Dobbs}, {Krumholz}, {Ballesteros-Paredes},
  {Bolatto}, {Fukui}, {Heyer}, {Low}, {Ostriker}, \&
  {V{\'a}zquez-Semadeni}}]{mc-formation-ppvi-review-2014}
{Dobbs}, C.~L., {Krumholz}, M.~R., {Ballesteros-Paredes}, J., {et~al.} 2014,
  Protostars and Planets VI, 1312.3223, 3

\bibitem[{{Dunham} {et~al.}(2014){Dunham}, {Arce}, {Mardones}, {Lee},
  {Matthews}, {Stutz}, \& {Williams}}]{outflows-in-low-mass-ps-Dunham-2014}
{Dunham}, M.~M., {Arce}, H.~G., {Mardones}, D., {et~al.} 2014, ApJ, 783, 29

\bibitem[{{Federrath} \&
  {Klessen}(2012)}]{pdf-simulations-turbulent-clouds-federrath-klessen-2012}
{Federrath}, C. \& {Klessen}, R.~S. 2012, ApJ, 761, 156

\bibitem[{{Federrath} \&
  {Klessen}(2013)}]{sf-efficiency-in-turbulent-magnetized-clouds-federrath-klessen-2013}
{Federrath}, C. \& {Klessen}, R.~S. 2013, ApJ, 763, 51

\bibitem[{{Federrath} {et~al.}(2010){Federrath}, {Roman-Duval}, {Klessen},
  {Schmidt}, \& {Mac Low}}]{turbulence-ism-forcing-federrath-2010}
{Federrath}, C., {Roman-Duval}, J., {Klessen}, R.~S., {Schmidt}, W., \& {Mac
  Low}, M.-M. 2010, A\&A, 512, A81

\bibitem[{{Girichidis} {et~al.}(2014){Girichidis}, {Konstandin}, {Whitworth},
  \& {Klessen}}]{pdf-column-density-evolution-numerical-girichidis-2014}
{Girichidis}, P., {Konstandin}, L., {Whitworth}, A.~P., \& {Klessen}, R.~S.
  2014, ApJ, 781, 91

\bibitem[{{G{\'o}mez} \&
  {V{\'a}zquez-Semadeni}(2014)}]{filaments-in-simulations-mc-formation-gomez-vazquez-semadeni-2014}
{G{\'o}mez}, G.~C. \& {V{\'a}zquez-Semadeni}, E. 2014, ApJ, 791, 124

\bibitem[{{Griffin} {et~al.}(2010){Griffin}, {Abergel}, {Abreu}, {Ade},
  {Andr{\'e}}, {Augueres}, {Babbedge}, {Bae}, {Baillie}, {Baluteau}, {Barlow},
  {Bendo}, {Benielli}, {Bock}, {Bonhomme}, {Brisbin}, {Brockley-Blatt},
  {Caldwell}, {Cara}, {Castro-Rodriguez}, {Cerulli}, {Chanial}, {Chen},
  {Clark}, {Clements}, {Clerc}, {Coker}, {Communal}, {Conversi}, {Cox},
  {Crumb}, {Cunningham}, {Daly}, {Davis}, {de Antoni}, {Delderfield}, {Devin},
  {di Giorgio}, {Didschuns}, {Dohlen}, {Donati}, {Dowell}, {Dowell}, {Duband},
  {Dumaye}, {Emery}, {Ferlet}, {Ferrand}, {Fontignie}, {Fox}, {Franceschini},
  {Frerking}, {Fulton}, {Garcia}, {Gastaud}, {Gear}, {Glenn}, {Goizel},
  {Griffin}, {Grundy}, {Guest}, {Guillemet}, {Hargrave}, {Harwit}, {Hastings},
  {Hatziminaoglou}, {Herman}, {Hinde}, {Hristov}, {Huang}, {Imhof}, {Isaak},
  {Israelsson}, {Ivison}, {Jennings}, {Kiernan}, {King}, {Lange}, {Latter},
  {Laurent}, {Laurent}, {Leeks}, {Lellouch}, {Levenson}, {Li}, {Li},
  {Lilienthal}, {Lim}, {Liu}, {Lu}, {Madden}, {Mainetti}, {Marliani}, {McKay},
  {Mercier}, {Molinari}, {Morris}, {Moseley}, {Mulder}, {Mur}, {Naylor},
  {Nguyen}, {O'Halloran}, {Oliver}, {Olofsson}, {Olofsson}, {Orfei}, {Page},
  {Pain}, {Panuzzo}, {Papageorgiou}, {Parks}, {Parr-Burman}, {Pearce},
  {Pearson}, {P{\'e}rez-Fournon}, {Pinsard}, {Pisano}, {Podosek}, {Pohlen},
  {Polehampton}, {Pouliquen}, {Rigopoulou}, {Rizzo}, {Roseboom}, {Roussel},
  {Rowan-Robinson}, {Rownd}, {Saraceno}, {Sauvage}, {Savage}, {Savini},
  {Sawyer}, {Scharmberg}, {Schmitt}, {Schneider}, {Schulz}, {Schwartz},
  {Shafer}, {Shupe}, {Sibthorpe}, {Sidher}, {Smith}, {Smith}, {Smith},
  {Spencer}, {Stobie}, {Sudiwala}, {Sukhatme}, {Surace}, {Stevens}, {Swinyard},
  {Trichas}, {Tourette}, {Triou}, {Tseng}, {Tucker}, {Turner}, {Vaccari},
  {Valtchanov}, {Vigroux}, {Virique}, {Voellmer}, {Walker}, {Ward}, {Waskett},
  {Weilert}, {Wesson}, {White}, {Whitehouse}, {Wilson}, {Winter}, {Woodcraft},
  {Wright}, {Xu}, {Zavagno}, {Zemcov}, {Zhang}, \& {Zonca}}]{spire}
{Griffin}, M.~J., {Abergel}, A., {Abreu}, A., {et~al.} 2010, A\&A, 518, L3

\bibitem[{{G{\"u}sten} {et~al.}(2006){G{\"u}sten}, {Nyman}, {Schilke},
  {Menten}, {Cesarsky}, \& {Booth}}]{apex}
{G{\"u}sten}, R., {Nyman}, L.~{\AA}., {Schilke}, P., {et~al.} 2006, A\&A, 454,
  L13

\bibitem[{{Hara} {et~al.}(1999){Hara}, {Tachihara}, {Mizuno}, {Onishi},
  {Kawamura}, {Obayashi}, \& {Fukui}}]{lupus-clouds-c18o-Hara-1999}
{Hara}, A., {Tachihara}, K., {Mizuno}, A., {et~al.} 1999, PASJ, 51, 895

\bibitem[{{Hartmann} {et~al.}(2001){Hartmann}, {Ballesteros-Paredes}, \&
  {Bergin}}]{rapid-formation-mcs-hartmann-2001}
{Hartmann}, L., {Ballesteros-Paredes}, J., \& {Bergin}, E.~A. 2001, ApJ, 562,
  852

\bibitem[{{Harvey} {et~al.}(2013){Harvey}, {Fallscheer}, {Ginsburg}, {Terebey},
  {Andr{\'e}}, {Bourke}, {Di Francesco}, {K{\"o}nyves}, {Matthews}, \&
  {Peterson}}]{pdf-auriga-california-herschel-harvey-2013}
{Harvey}, P.~M., {Fallscheer}, C., {Ginsburg}, A., {et~al.} 2013, ApJ, 764, 133

\bibitem[{{Juvela} {et~al.}(2013){Juvela}, {Malinen}, \&
  {Lunttila}}]{column-density-maps-herschel-juvela-2013}
{Juvela}, M., {Malinen}, J., \& {Lunttila}, T. 2013, A\&A, 553, A113

\bibitem[{{Kainulainen} {et~al.}(2011){Kainulainen}, {Beuther}, {Banerjee},
  {Federrath}, \& {Henning}}]{pdf-column-density-kainulainen-2011}
{Kainulainen}, J., {Beuther}, H., {Banerjee}, R., {Federrath}, C., \&
  {Henning}, T. 2011, A\&A, 530, A64

\bibitem[{{Kainulainen} {et~al.}(2009){Kainulainen}, {Beuther}, {Henning}, \&
  {Plume}}]{pdf-column-density-kainulainen-2009}
{Kainulainen}, J., {Beuther}, H., {Henning}, T., \& {Plume}, R. 2009, A\&A,
  508, L35

\bibitem[{{Kalberla} {et~al.}(2010){Kalberla}, {McClure-Griffiths}, {Pisano},
  {Calabretta}, {Ford}, {Lockman}, {Staveley-Smith}, {Kerp}, {Winkel},
  {Murphy}, \& {Newton-McGee}}]{gass-hi-kalberla-2010}
{Kalberla}, P.~M.~W., {McClure-Griffiths}, N.~M., {Pisano}, D.~J., {et~al.}
  2010, A\&A, 521, A17

\bibitem[{{Kevlahan} \& {Pudritz}(2009)}]{shocks-ism-imf-kevlahan-2009}
{Kevlahan}, N. \& {Pudritz}, R.~E. 2009, ApJ, 702, 39

\bibitem[{{Klessen} {et~al.}(2000){Klessen}, {Heitsch}, \& {Mac
  Low}}]{gravitational-collaps-in-mc-klessen-2000}
{Klessen}, R.~S., {Heitsch}, F., \& {Mac Low}, M.-M. 2000, ApJ, 535, 887

\bibitem[{{Krause} {et~al.}(2014){Krause}, {Diehl}, {B{\"o}hringer},
  {Freyberg}, \& {Lubos}}]{feedback-superbubbles-xrays-krause-2014}
{Krause}, M., {Diehl}, R., {B{\"o}hringer}, H., {Freyberg}, M., \& {Lubos}, D.
  2014, A\&A, 566, A94

\bibitem[{{Krause} {et~al.}(2013){Krause}, {Fierlinger}, {Diehl}, {Burkert},
  {Voss}, \& {Ziegler}}]{superbubbles-vishniac-krause-2013}
{Krause}, M., {Fierlinger}, K., {Diehl}, R., {et~al.} 2013, A\&A, 550, A49

\bibitem[{{Lombardi} {et~al.}(2014){Lombardi}, {Bouy}, {Alves}, \&
  {Lada}}]{herschel-planck-column-density-maps-lombardi-2014}
{Lombardi}, M., {Bouy}, H., {Alves}, J., \& {Lada}, C.~J. 2014, A\&A, 566, A45

\bibitem[{{Lombardi} {et~al.}(2008){Lombardi}, {Lada}, \&
  {Alves}}]{lupus-2mass-extinction-maps-lombardi-2008}
{Lombardi}, M., {Lada}, C.~J., \& {Alves}, J. 2008, A\&A, 489, 143

\bibitem[{{Mathis} {et~al.}(1977){Mathis}, {Rumpl}, \&
  {Nordsieck}}]{dust-opacity-mrn-1977}
{Mathis}, J.~S., {Rumpl}, W., \& {Nordsieck}, K.~H. 1977, ApJ, 217, 425

\bibitem[{{Matsumoto} {et~al.}(2015){Matsumoto}, {Dobashi}, \&
  {Shimoikura}}]{pdf-double-peaks-colliding-flows-matsumoto-2015}
{Matsumoto}, T., {Dobashi}, K., \& {Shimoikura}, T. 2015, ApJ, 801, 77

\bibitem[{{Matthews} {et~al.}(2014){Matthews}, {Ade}, {Angil{\`e}}, {Benton},
  {Chapin}, {Chapman}, {Devlin}, {Fissel}, {Fukui}, {Gandilo}, {Gundersen},
  {Hargrave}, {Klein}, {Korotkov}, {Moncelsi}, {Mroczkowski}, {Netterfield},
  {Novak}, {Nutter}, {Olmi}, {Pascale}, {Poidevin}, {Savini}, {Scott},
  {Shariff}, {Soler}, {Tachihara}, {Thomas}, {Truch}, {Tucker}, {Tucker}, \&
  {Ward-Thompson}}]{lupus-submm-polarimetry-matthews-2014}
{Matthews}, T.~G., {Ade}, P.~A.~R., {Angil{\`e}}, F.~E., {et~al.} 2014, ApJ,
  784, 116

\bibitem[{{McClure-Griffiths} {et~al.}(2009){McClure-Griffiths}, {Pisano},
  {Calabretta}, {Ford}, {Lockman}, {Staveley-Smith}, {Kalberla}, {Bailin},
  {Dedes}, {Janowiecki}, {Gibson}, {Murphy}, {Nakanishi}, \&
  {Newton-McGee}}]{gass-hi-mcclure-griffiths-2009}
{McClure-Griffiths}, N.~M., {Pisano}, D.~J., {Calabretta}, M.~R., {et~al.}
  2009, ApJS, 181, 398

\bibitem[{{Mer{\'{\i}}n} {et~al.}(2008){Mer{\'{\i}}n}, {J{\o}rgensen},
  {Spezzi}, {Alcal{\'a}}, {Evans}, {Harvey}, {Prusti}, {Chapman}, {Huard}, {van
  Dishoeck}, \& {Comer{\'o}n}}]{lupus-clouds-spitzer-c2d-merin-2008}
{Mer{\'{\i}}n}, B., {J{\o}rgensen}, J., {Spezzi}, L., {et~al.} 2008, ApJS, 177,
  551

\bibitem[{{Ossenkopf} \&
  {Henning}(1994)}]{dust-opacities-for-ps-cores-ossenkopf-1994}
{Ossenkopf}, V. \& {Henning}, T. 1994, A\&A, 291, 943

\bibitem[{{Ott}(2010)}]{hipe}
{Ott}, S. 2010, in Astronomical Society of the Pacific Conference Series, Vol.
  434, Astronomical Data Analysis Software and Systems XIX, ed. {Y.~Mizumoto,
  K.-I.~Morita, \& M.~Ohishi}, 139

\bibitem[{{Oya} {et~al.}(2014){Oya}, {Sakai}, {Sakai}, {Watanabe}, {Hirota},
  {Lindberg}, {Bisschop}, {J{\o}rgensen}, {van Dishoeck}, \&
  {Yamamoto}}]{lupusI-iras-outflow-oya-2014}
{Oya}, Y., {Sakai}, N., {Sakai}, T., {et~al.} 2014, ApJ, 795, 152

\bibitem[{{Palmeirim} {et~al.}(2013){Palmeirim}, {Andr{\'e}}, {Kirk},
  {Ward-Thompson}, {Arzoumanian}, {K{\"o}nyves}, {Didelon}, {Schneider},
  {Benedettini}, {Bontemps}, {Di Francesco}, {Elia}, {Griffin}, {Hennemann},
  {Hill}, {Martin}, {Men'shchikov}, {Molinari}, {Motte}, {Nguyen Luong},
  {Nutter}, {Peretto}, {Pezzuto}, {Roy}, {Rygl}, {Spinoglio}, \&
  {White}}]{herschel-column-density-pdf-taurus-palmeirim-2013}
{Palmeirim}, P., {Andr{\'e}}, P., {Kirk}, J., {et~al.} 2013, A\&A, 550, A38

\bibitem[{{Pilbratt} {et~al.}(2010){Pilbratt}, {Riedinger}, {Passvogel},
  {Crone}, {Doyle}, {Gageur}, {Heras}, {Jewell}, {Metcalfe}, {Ott}, \&
  {Schmidt}}]{herschel}
{Pilbratt}, G.~L., {Riedinger}, J.~R., {Passvogel}, T., {et~al.} 2010, A\&A,
  518, L1

\bibitem[{{Planck Collaboration} {et~al.}(2014){Planck Collaboration},
  {Abergel}, {Ade}, {Aghanim}, {Alves}, {Aniano}, {Armitage-Caplan}, {Arnaud},
  {Ashdown}, {Atrio-Barandela}, \&
  et~al.}]{planck-2013-sky-thermal-dust-emission-model}
{Planck Collaboration}, {Abergel}, A., {Ade}, P.~A.~R., {et~al.} 2014, A\&A,
  571, A11

\bibitem[{{Poglitsch} {et~al.}(2010){Poglitsch}, {Waelkens}, {Geis},
  {Feuchtgruber}, {Vandenbussche}, {Rodriguez}, {Krause}, {Renotte}, {van
  Hoof}, {Saraceno}, {Cepa}, {Kerschbaum}, {Agn{\`e}se}, {Ali}, {Altieri},
  {Andreani}, {Augueres}, {Balog}, {Barl}, {Bauer}, {Belbachir}, {Benedettini},
  {Billot}, {Boulade}, {Bischof}, {Blommaert}, {Callut}, {Cara}, {Cerulli},
  {Cesarsky}, {Contursi}, {Creten}, {De Meester}, {Doublier}, {Doumayrou},
  {Duband}, {Exter}, {Genzel}, {Gillis}, {Gr{\"o}zinger}, {Henning},
  {Herreros}, {Huygen}, {Inguscio}, {Jakob}, {Jamar}, {Jean}, {de Jong},
  {Katterloher}, {Kiss}, {Klaas}, {Lemke}, {Lutz}, {Madden}, {Marquet},
  {Martignac}, {Mazy}, {Merken}, {Montfort}, {Morbidelli}, {M{\"u}ller},
  {Nielbock}, {Okumura}, {Orfei}, {Ottensamer}, {Pezzuto}, {Popesso},
  {Putzeys}, {Regibo}, {Reveret}, {Royer}, {Sauvage}, {Schreiber}, {Stegmaier},
  {Schmitt}, {Schubert}, {Sturm}, {Thiel}, {Tofani}, {Vavrek}, {Wetzstein},
  {Wieprecht}, \& {Wiezorrek}}]{pacs}
{Poglitsch}, A., {Waelkens}, C., {Geis}, N., {et~al.} 2010, A\&A, 518, L2

\bibitem[{{Preibisch} {et~al.}(2002){Preibisch}, {Brown}, {Bridges},
  {Guenther}, \& {Zinnecker}}]{usco-full-population-preibisch-2002}
{Preibisch}, T., {Brown}, A.~G.~A., {Bridges}, T., {Guenther}, E., \&
  {Zinnecker}, H. 2002, AJ, 124, 404

\bibitem[{{Preibisch} \& {Mamajek}(2008)}]{sco-cen-sfhb-preibisch-mamajek-2008}
{Preibisch}, T. \& {Mamajek}, E. 2008, Handbook of Star Forming Regions, Volume
  II, ed. B.~{Reipurth}, 235

\bibitem[{{Preibisch} \&
  {Zinnecker}(2007)}]{ob-stars-triggered-sf-preibisch-zinnecker-2007}
{Preibisch}, T. \& {Zinnecker}, H. 2007, in IAU Symposium, Vol. 237, IAU
  Symposium, ed. {B.~G.~Elmegreen \& J.~Palous}, 270--277

\bibitem[{{Roccatagliata} {et~al.}(2015){Roccatagliata}, {Dale}, {Ratzka},
  {Testi}, {Burkert}, {Koepferl}, {Sicilia-Aguilar}, {Eiroa}, \&
  {Gaczkowski}}]{pdf-serpens-core-roccatagliata-2015}
{Roccatagliata}, V., {Dale}, J.~E., {Ratzka}, T., {et~al.} 2015, A\&A,
  submitted

\bibitem[{{Roussel}(2013)}]{scanamorphos}
{Roussel}, H. 2013, PASP, 125, 1126

\bibitem[{{Rygl} {et~al.}(2013){Rygl}, {Benedettini}, {Schisano}, {Elia},
  {Molinari}, {Pezzuto}, {Andr{\'e}}, {Bernard}, {White}, {Polychroni},
  {Bontemps}, {Cox}, {Di Francesco}, {Facchini}, {Fallscheer}, {di Giorgio},
  {Hennemann}, {Hill}, {K{\"o}nyves}, {Minier}, {Motte}, {Nguyen-Luong},
  {Peretto}, {Pestalozzi}, {Sadavoy}, {Schneider}, {Spinoglio}, {Testi}, \&
  {Ward-Thompson}}]{lupus-clouds-herschel-rygl-2013}
{Rygl}, K.~L.~J., {Benedettini}, M., {Schisano}, E., {et~al.} 2013, A\&A, 549,
  L1

\bibitem[{{Schneider} {et~al.}(2013){Schneider}, {Andr{\'e}}, {K{\"o}nyves},
  {Bontemps}, {Motte}, {Federrath}, {Ward-Thompson}, {Arzoumanian},
  {Benedettini}, {Bressert}, {Didelon}, {Di Francesco}, {Griffin}, {Hennemann},
  {Hill}, {Palmeirim}, {Pezzuto}, {Peretto}, {Roy}, {Rygl}, {Spinoglio}, \&
  {White}}]{pdf-column-density-orion-schneider-2013}
{Schneider}, N., {Andr{\'e}}, P., {K{\"o}nyves}, V., {et~al.} 2013, ApJL, 766,
  L17

\bibitem[{{Schneider} {et~al.}(2012){Schneider}, {Csengeri}, {Hennemann},
  {Motte}, {Didelon}, {Federrath}, {Bontemps}, {Di Francesco}, {Arzoumanian},
  {Minier}, {Andr{\'e}}, {Hill}, {Zavagno}, {Nguyen-Luong}, {Attard},
  {Bernard}, {Elia}, {Fallscheer}, {Griffin}, {Kirk}, {Klessen}, {K{\"o}nyves},
  {Martin}, {Men'shchikov}, {Palmeirim}, {Peretto}, {Pestalozzi}, {Russeil},
  {Sadavoy}, {Sousbie}, {Testi}, {Tremblin}, {Ward-Thompson}, \&
  {White}}]{pdf-rosette-cluster-formation-Schneider-2012}
{Schneider}, N., {Csengeri}, T., {Hennemann}, M., {et~al.} 2012, A\&A, 540, L11

\bibitem[{{Schneider} {et~al.}(2015){Schneider}, {Ossenkopf}, {Csengeri},
  {Klessen}, {Federrath}, {Tremblin}, {Girichidis}, {Bontemps}, \&
  {Andr{\'e}}}]{pdf-column-density-schneider-2015}
{Schneider}, N., {Ossenkopf}, V., {Csengeri}, T., {et~al.} 2015, A\&A, 575, A79

\bibitem[{{Schuller}(2012)}]{boa}
{Schuller}, F. 2012, in Society of Photo-Optical Instrumentation Engineers
  (SPIE) Conference Series, Vol. 8452, Society of Photo-Optical Instrumentation
  Engineers (SPIE) Conference Series

\bibitem[{{Schuller} {et~al.}(2009){Schuller}, {Menten}, {Contreras},
  {Wyrowski}, {Schilke}, {Bronfman}, {Henning}, {Walmsley}, {Beuther},
  {Bontemps}, {Cesaroni}, {Deharveng}, {Garay}, {Herpin}, {Lefloch}, {Linz},
  {Mardones}, {Minier}, {Molinari}, {Motte}, {Nyman}, {Reveret}, {Risacher},
  {Russeil}, {Schneider}, {Testi}, {Troost}, {Vasyunina}, {Wienen}, {Zavagno},
  {Kovacs}, {Kreysa}, {Siringo}, \& {Wei{\ss}}}]{atlasgal}
{Schuller}, F., {Menten}, K.~M., {Contreras}, Y., {et~al.} 2009, A\&A, 504, 415

\bibitem[{{Siringo} {et~al.}(2009){Siringo}, {Kreysa}, {Kov{\'a}cs},
  {Schuller}, {Wei{\ss}}, {Esch}, {Gem{\"u}nd}, {Jethava}, {Lundershausen},
  {Colin}, {G{\"u}sten}, {Menten}, {Beelen}, {Bertoldi}, {Beeman}, \&
  {Haller}}]{laboca}
{Siringo}, G., {Kreysa}, E., {Kov{\'a}cs}, A., {et~al.} 2009, A\&A, 497, 945

\bibitem[{{Tachihara} {et~al.}(1996){Tachihara}, {Dobashi}, {Mizuno}, {Ogawa},
  \& {Fukui}}]{lupus-clouds-13co(1-0)-Tachihara-1996}
{Tachihara}, K., {Dobashi}, K., {Mizuno}, A., {Ogawa}, H., \& {Fukui}, Y. 1996,
  PASJ, 48, 489

\bibitem[{{Tachihara} {et~al.}(2001){Tachihara}, {Toyoda}, {Onishi}, {Mizuno},
  {Fukui}, \& {Neuh{\"a}user}}]{lupus-clouds-12co-Tachihara-2001}
{Tachihara}, K., {Toyoda}, S., {Onishi}, T., {et~al.} 2001, PASJ, 53, 1081

\bibitem[{{Tothill} {et~al.}(2009){Tothill}, {L{\"o}hr}, {Parshley}, {Stark},
  {Lane}, {Harnett}, {Wright}, {Walker}, {Bourke}, \&
  {Myers}}]{lupus-clouds-13co(2-1)-Tothill-2009}
{Tothill}, N.~F.~H., {L{\"o}hr}, A., {Parshley}, S.~C., {et~al.} 2009, ApJS,
  185, 98

\bibitem[{{Tremblin} {et~al.}(2012){Tremblin}, {Audit}, {Minier}, {Schmidt}, \&
  {Schneider}}]{pdf-hii-turbulence-shock-simulation-tremblin-2012}
{Tremblin}, P., {Audit}, E., {Minier}, V., {Schmidt}, W., \& {Schneider}, N.
  2012, A\&A, 546, A33

\bibitem[{{Tremblin} {et~al.}(2014){Tremblin}, {Schneider}, {Minier},
  {Didelon}, {Hill}, {Anderson}, {Motte}, {Zavagno}, {Andr{\'e}},
  {Arzoumanian}, {Audit}, {Benedettini}, {Bontemps}, {Csengeri}, {Di
  Francesco}, {Giannini}, {Hennemann}, {Nguyen Luong}, {Marston}, {Peretto},
  {Rivera-Ingraham}, {Russeil}, {Rygl}, {Spinoglio}, \&
  {White}}]{herschel-hii-compression-pdf-tremblin-2014}
{Tremblin}, P., {Schneider}, N., {Minier}, V., {et~al.} 2014, A\&A, 564, A106

\bibitem[{{Vazquez-Semadeni}(1994)}]{structure-flows-ism-vazquez-1994}
{Vazquez-Semadeni}, E. 1994, ApJ, 423, 681

\bibitem[{{V{\'a}zquez-Semadeni} {et~al.}(2011){V{\'a}zquez-Semadeni},
  {Banerjee}, {G{\'o}mez}, {Hennebelle}, {Duffin}, \&
  {Klessen}}]{mc-formation-mhd-vazquez-semadeni-2011}
{V{\'a}zquez-Semadeni}, E., {Banerjee}, R., {G{\'o}mez}, G.~C., {et~al.} 2011,
  MNRAS, 414, 2511

\bibitem[{{V{\'a}zquez-Semadeni} {et~al.}(2007){V{\'a}zquez-Semadeni},
  {G{\'o}mez}, {Jappsen}, {Ballesteros-Paredes}, {Gonz{\'a}lez}, \&
  {Klessen}}]{mc-formation-stars-vazquez-semadeni-2007}
{V{\'a}zquez-Semadeni}, E., {G{\'o}mez}, G.~C., {Jappsen}, A.~K., {et~al.}
  2007, ApJ, 657, 870

\bibitem[{{Vilas-Boas} {et~al.}(2000){Vilas-Boas}, {Myers}, \&
  {Fuller}}]{lupus-clouds-13co-cores-vilas-boas-2000}
{Vilas-Boas}, J.~W.~S., {Myers}, P.~C., \& {Fuller}, G.~A. 2000, ApJ, 532, 1038

\bibitem[{{Ward} {et~al.}(2014){Ward}, {Wadsley}, \&
  {Sills}}]{pdf-column-density-simu-ward-2014}
{Ward}, R.~L., {Wadsley}, J., \& {Sills}, A. 2014, MNRAS, 445, 1575

\bibitem[{{Williams} {et~al.}(1994){Williams}, {de Geus}, \&
  {Blitz}}]{clumpfind}
{Williams}, J.~P., {de Geus}, E.~J., \& {Blitz}, L. 1994, ApJ, 428, 693

\end{thebibliography}

\end{document}